\documentclass[12pt]{article}

\usepackage{a4wide}
\usepackage{amsmath}
\usepackage{amsfonts}
\usepackage{amsthm}

\newcommand{\reals}{\mathbf{R}}%    real line
\newcommand{\sphere}{\mathbf{S}}%   sphere
\newcommand{\torus}{\mathbf{T}}%    torus
\newcommand{\lens}{\mathbf{L}}%     lens space

\newcommand{\U}{\mathcal{U}}

\newcommand{\z}{\mathbf{z}}
\newcommand{\A}{\mathbf{A}}
\newcommand{\N}{\mathbf{N}}
\newcommand{\f}{\mathbf{f}}
\newcommand{\vv}{\mathbf{v}}
\newcommand{\g}{\mathbf{g}}
\newcommand{\Ubar}{\overline{U}}% vector U
\newcommand{\Vbar}{\overline{V}}% vector V

\newcommand{\Ordo}{\mathcal{O}}%       ordo
\newcommand{\D}{\mathcal{D}}%          covariant derivative of t-\theta metric
\newcommand{\grad}{\nabla}%            gradient
\newcommand{\lap}{\Delta}%             Laplacian
\newcommand{\id}{\mathrm{d}}%          integral d
\renewcommand{\d}{\partial}%           partial d (overriding \d accent)
\newcommand{\dd}[1]{\frac{\d}{\d #1}}% derivative vector
\newcommand{\tr}{\operatorname{tr}}%   trace
\newcommand{\closure}{\overline}

\newcommand{\abs}[1]{\lvert#1\rvert}%               absolute value
\newcommand{\norm}[1]{\lVert{#1}\rVert}%            basic norm
\newcommand{\normLinf}[1]{\norm{#1}_{L^\infty}}%    L^oo norm
\newcommand{\normHs}[1]{\norm{#1}_{H^s}}%           H^s norm
\newcommand{\iprod}[2]{\langle{#1},{#2}\rangle}%    inner product
\newcommand{\iprodLii}[2]{\iprod{#1}{#2}_{L^2}}%    L^2 inner product
\newtheorem{theorem}{Theorem}[section]
\newtheorem{lemma}{Lemma}[section]
\theoremstyle{definition}
\newtheorem{definition}{Definition}[section]

\newcommand{\ignorewidth}[1]{\rlap{$\displaystyle #1$}}

\newcommand{\updatebox}{\fbox{\small\sffamily\bfseries Update}}
\newcommand{\update}{\marginpar{\updatebox}}
  {\begin{sffamily}\small\update\begin{itemize}}%
  {\end{itemize}\end{sffamily}}

\begin{document}

\title{Fuchsian analysis of $\sphere^2\times\sphere^1$
       and $\sphere^3$ Gowdy spacetimes}
\author{Fredrik St{\aa}hl%
        \thanks{E-mail address: fredrik@aei-potsdam.mpg.de}\\
        Albert-Einstein-Institut,\\
        Am M\"uhlenberg~1, 14476 Golm, Germany\\
        Report no. AEI-2001-110}
\date{September 4, 2001}

\maketitle

\begin{abstract}
  The Gowdy spacetimes are vacuum solutions of Einstein's equations
  with two commuting Killing vectors having compact spacelike orbits
  with $\torus^3$, $\sphere^2\times\sphere^1$ or $\sphere^3$ topology.
  In the case of $\torus^3$ topology, Kichenassamy and Rendall have
  found a family of singular solutions which are asymptotically
  velocity dominated by construction.  In the case when the velocity
  is between zero and one, the solutions depend on the maximal number
  of free functions.  We consider the similar case with
  $\sphere^2\times\sphere^1$ or $\sphere^3$ topology, where the main
  complication is the presence of symmetry axes.  We use Fuchsian
  techniques to show the existence of singular solutions similar to
  the $\torus^3$ case.  We first solve the analytic case and then
  generalise to the smooth case by approximating smooth data with a
  sequence of analytic data.  However, for the metric to be smooth at
  the axes, the velocity must be $1$ or $3$ there, which is outside
  the range where the constructed solutions depend on the full number
  of free functions.  A plausible explanation is that in general a
  spiky feature may develop at the axis, a situation which is
  unsuitable for a direct treatment by Fuchsian methods.
\end{abstract}

%%%%%%%%%%%%%%%%%%%%%%%%%%%%%%%%%%%%%%%%%%%%%%%%%%%%%%%%%%%%%%%%%%%%%%%%%%%%%%%

\section{Introduction}
\label{sec:intro}

The singularity theorems by Hawking and Penrose show that in general
spacetimes, singularities will be present \cite{Hawking-Ellis}.  Apart
from existence, these results give very little information on the
nature of these singularities.

There is however a general proposal for the structure of the generic
cosmological singularity, put forth by Belinskii, Khalatnikov and
Lifshitz (BKL) \cite{BKL:general-time-sing}.  The argument is based on
formal calculations which originally were without rigorous
mathematical proofs.  In essence, the BKL picture consists of two
predictions: firstly, that the evolution of a spatial point near the
singularity is unaffected by the evolution of nearby points.  Thus the
dynamics of an inhomogeneous model can be described by a homogeneous
model at each spatial point, with the parameters of the homogeneous
model depending on the spatial coordinates.  Secondly, the dynamics
are expected be oscillatory, similar to the situation in the
homogeneous Bianchi IX models.  Loosely speaking, the evolution of a
spatial point near a cosmological singularity should appear as an
infinite sequence of epochs, each similar to a different Kasner
spacetime.

Some cosmological models exhibit a simpler behaviour without
oscillations, termed `asymptotically velocity dominated' (AVD)
\cite{Eardley-Liang-Sachs:VTD-dust,Isenberg-Moncrief:Gowdy-polar-asymptotic}.
For these models there is no oscillation and the evolution of a given
spatial point approaches a particular Kasner solution. This is
exhibited in the evolution equations by the spatial derivative terms
becoming negligible in relation to time derivatives.

The behaviour of general spacetimes is still out of reach of the
mathematical techniques available, so it makes sense to study
particular subclasses of solutions, subject to some particular
symmetry and/or a particular choice of matter.  Recently there has
been considerable progress in the study of spatially homogeneous
spacetimes \cite{Rendall-Tod:EV-LRS,Rendall-Uggla:EV-LRS,%
Ringstrom:curvature-blow-up,Ringstrom:BianchiIX,%
Weaver:magn-Bianchi-IX0}.  There is also an interesting result which
establishes AVD behaviour for the general Einstein equations coupled
to a scalar field, without any symmetry assumptions
\cite{Andersson-Rendall:quiescent-sings}.

It is of obvious interest to relax the symmetry assumptions from
homogeneous to inhomogeneous spacetimes.  For simplicity, we restrict
attention to the spatially compact case in vacuum.  The largest
symmetry group that is compatible with inhomogeneities is then
$U(1)\times U(1)$, which in the spatially compact case leads to the
Gowdy models \cite{Gowdy:Gowdy,Chrusciel:U1xU1}.  The possible spatial
topologies are then essentially $\torus^3$, $\sphere^2\times\sphere^1$
or $\sphere^3$. (In the more general case of \emph{local} $U(1)\times
U(1)$ symmetry, other topologies are possible, see
\cite{Tanimoto:local-U1xU1}.)

For Gowdy spacetimes it is possible to define a quantity, the
`velocity', which describes the dynamics near the singularity. The
terminology is slightly misleading since the velocity is only uniquely
defined up to a sign. While a sign can be introduced, it depends on
the particular parametrisation of the metric. If AVD holds, the
velocity has a (space dependent) limit at the singularity.

For the subclass of polarised Gowdy models, which describe universa
with polarised gravitational waves, the Einstein evolution equations
reduce to a single linear ordinary differential equation, which can be
solved formally in terms of a series of Legendre polynomials.  The
polarised solutions have been shown to be AVD
\cite{Chrusciel-Isenberg-Moncrief:Gowdy-SCC,%
Isenberg-Moncrief:Gowdy-polar-asymptotic,Hanquin-Demaret:Gowdy-S2xS1+S3}.

For the full Gowdy models with $\torus^3$ spatial topology, numerical
simulations show that the velocity has a limit between $0$ and $1$ for
almost all spatial points in generic situations
\cite{Berger-Moncrief:Gowdy-T3-num,Berger-Garfinkle:Gowdy-T3-phenom,%
Hern-Stewart,Hern:num-inhom}.
The exceptions are isolated points where the velocity remains outside
this range, which results in a discontinuous behaviour, `spikes', in
the limit.  A family of singular solutions have been constructed
analytically by Kichenassamy and Rendall using Fuchsian methods
\cite{Kichen-Rendall:Gowdy-T3-analytic}, based on previous work on
formal asymptotic expansions by Grubisic and Moncrief
\cite{Grubisic-Moncrief:Gowdy-T3-asymptotic}.  These solutions have a
prescribed asymptotic behaviour.  In the case when the asymptotic
velocity is between $0$ and $1$, the solutions depend on the full
number of free functions, which indicates that they correspond to an
open set of initial data.  Recently, a similar family of solutions
with spikes have been constructed by Rendall and Weaver, showing that
the spikes are real features of the model and not numerical artefacts
\cite{Rendall-Weaver:manufacture}.  We should also mention that there
is a result on the existence of Gowdy $\torus^3$ solutions with small
data close to that of a Kasner $(2/3,2/3,-1/3)$ spacetime
\cite{Chrusciel:uniqueness-II}.

For the unpolarised Gowdy spacetimes with $\sphere^2\times\sphere^1$
or $\sphere^3$ topology, much less is known.  Garfinkle has done
numerical simulations of Gowdy $\sphere^2\times\sphere^1$ spacetimes,
which show similar behaviour as in the $\torus^3$ case, including the
appearance of spikes.  It is the purpose of this paper to provide a
Fuchsian analysis of Gowdy $\sphere^2\times\sphere^1$ and $\sphere^3$
spacetimes similar to that in \cite{Kichen-Rendall:Gowdy-T3-analytic}
for $\torus^3$ spacetimes.

The Fuchsian algorithm
\cite{Kichenassamy:wave-eq,Kichen-Rendall:Gowdy-T3-analytic} has
proven to be a valuable tool in showing the existence of singular
solutions to nonlinear wave equations.  Variants of it has been used
to show the existence of Cauchy horizons with certain properties
\cite{Moncrief:U1-Cauchy-horizons}, to analyse the properties of
isotropic singularities
\cite{Newman:conformal-sing,Newman:conformal-sing-II,Claudel:Cauchy,%
Anguige-Tod:isotropic-I,Anguige-Tod:isotropic-II,Anguige:isotropic-III},
to construct families of singular Gowdy or plane symmetric spacetimes
\cite{Kichen-Rendall:Gowdy-T3-analytic,Anguige:plane-PF-Kasner,%
Anguige:Gowdy-PF-Kasner,Narita-Torii-Maeda:Gowdy-string,%
Isenberg-Kichenassamy:Gowdy-T2-polar}
and for the general result on scalar field spacetimes
\cite{Andersson-Rendall:quiescent-sings}.

The original version of the Fuchsian technique only applies to the
analytic case, since the proof uses the Cauchy formula in the complex
domain to estimate derivatives in terms of functions themselves.
There are a number of results extending the Fuchsian algorithm to give
smooth solutions
\cite{Newman:conformal-sing,Newman:conformal-sing-II,Claudel:Cauchy,%
Kichenassamy:Fuchsian-blow-up,Anguige:plane-PF-Kasner,%
Rendall:Gowdy-T3-smooth}
in various contexts.  In section~\ref{sec:smooth} below we will modify
the argument of \cite{Rendall:Gowdy-T3-smooth} to apply to our case.

Finally, we note that there is an additional motivation for studying
Gowdy spacetimes with $\sphere^2\times\sphere^1$ topology apart from
cosmology, since axisymmetric and stationary black hole interiors are
in fact Gowdy $\sphere^2\times\sphere^1$ spacetimes. These include the
region of Kerr spacetime between the inner and outer horizon
\cite{Obregon-Ryan:Gowdy-S2xS1-Kerr} and the interiors of the
`distorted black holes' studied by Geroch and Hartle
\cite{Geroch-Hartle:distorted-BH}.

The outline of the paper is as follows.  In section~\ref{sec:gowdy} we
give an introduction to the Gowdy spacetimes, introduce coordinates
and a parametrisation of the metric, and rewrite the field equations
in a suitable form.  We also discuss the restrictions needed for the
metric to be smooth at the axes and the properties of some previously
known solutions.  In section~\ref{sec:analytic} we give a short review
of the Fuchsian technique of \cite{Kichen-Rendall:Gowdy-T3-analytic},
and construct analytic solutions of the field equations.  This is then
extended to smooth solutions in section~\ref{sec:smooth}, by means of
a generalisation of the technique in \cite{Rendall:Gowdy-T3-smooth}.
Finally we discuss the shortcomings and implications of the results in
section~\ref{sec:discussion}.

%%%%%%%%%%%%%%%%%%%%%%%%%%%%%%%%%%%%%%%%%%%%%%%%%%%%%%%%%%%%%%%%%%%%%%%%%%%%%%%

\section{Gowdy spacetimes}
\label{sec:gowdy}

We assume that $(M,g)$ is a $U(1)\times U(1)$-symmetric spatially
compact spacetime with spacelike group orbits. More precisely, let
$M=I\times\Sigma$ where $I\subset\reals$ and $\Sigma$ is a connected
orientable compact three manifold, and assume that the $U(1)\times
U(1)$ group acts effectively on $\Sigma$. It then follows
\cite{Chrusciel:U1xU1} that the topology of $\Sigma$ is that of either
$\torus^3$, $\sphere^2\times\sphere^1$, $\sphere^3$ or a lens space
$\lens(p,q)$.  Since the $\torus^3$ case is treated in
\cite{Kichen-Rendall:Gowdy-T3-analytic} and
\cite{Rendall:Gowdy-T3-smooth} and $\lens(p,q)$ may be covered by
$\sphere^3$, we only consider $\Sigma\simeq\sphere^2\times\sphere^1$
or $\sphere^3$ here.%
\footnote{%
  \cite{Kichen-Rendall:Gowdy-T3-analytic} and
  \cite{Rendall:Gowdy-T3-smooth} treat the $\torus^3$ case with
  vanishing twist. See \cite{Isenberg-Kichenassamy:Gowdy-T2-polar} for
  a similar treatment with nonvanishing twist but restricted to the
  polarised case.
}%

We introduce coordinates on $\Sigma$ as follows.  Let $\theta$ label
the orbits of $U(1)\times U(1)$ in $\Sigma$ and let $(\phi,\chi)$ be
coordinates on each orbit induced by the standard coordinates on
$U(1)\times U(1)$.  For $\Sigma\simeq\sphere^2\times\sphere^1$, we
choose $\chi$ to be a cyclic coordinate on the $\sphere^1$ part,
$\chi\in[0,2\pi]\mod2\pi$, and $(\theta,\phi)$ to be spherical
coordinates on the $\sphere^2$ part, $\theta\in[0,\pi]$ and
$\phi\in[0,2\pi]\mod2\pi$.  This fixes $(\phi,\chi)$ up to
translation.  For $\Sigma\simeq\sphere^3$, the fix points to the first
$U(1)$ factor of $U(1)\times U(1)$ form a circle $S_1$, and the set of
fix points to the second factor is a similar circle $S_2$ (this is a
nontrivial consequence of the effective action, see, e.g.,
\cite{Mostert:compact-action}).  The parametrisation is chosen such
that $S_1$ and $S_2$ correspond to $\theta=0$ and $\theta=\pi$,
respectively.  This fixes $(\phi,\chi)$ up to translation in this case
as well.  Note that the two cases are similar in a neighbourhood of
one of the axes.

If $(M,g)$ is assumed to be a maximal globally hyperbolic development
of generic data on $\Sigma$, then $M$ contains the set
$(0,\pi)\times\Sigma$, on which the line element of the metric $g$ may
be parametrised as
\begin{equation}\label{eq:g-general}
  \id s^2 = e^{A}(-\id t^2 + \id\theta^2) + \id\sigma^2,
\end{equation}
where $\id\sigma^2$ is a metric on the orbits with determinant
$R=c\sin{t}\sin\theta$ for some positive constant $c$
\cite{Chrusciel:U1xU1}.  We set $c=1$ from now on, since a constant
conformal factor does not affect the qualitative aspects of the model.
Here `generic' refers to an open and dense subset of all Cauchy data;
see \cite{Chrusciel:U1xU1} for the exact definition.  We will
call this model the Gowdy model over $\Sigma$ \cite{Gowdy:Gowdy}.

In the $\torus^3$ case, the metric can also be written in the form
\eqref{eq:g-general}, but with orbit metric determinant $t$. It
follows that for $\sphere^2\times\sphere^1$ or $\sphere^3$, the metric
can be brought into the $\torus^3$ form by a change of coordinates
whenever $\theta\ne0$ and $\theta\ne\pi$. Thus the result of
\cite{Kichen-Rendall:Gowdy-T3-analytic} and
\cite{Rendall:Gowdy-T3-smooth} may be applied locally on any subset of
$M$ where $\theta$ is bounded away from $0$ and $\pi$, and what
remains is to study a neighbourhood of one of the symmetry axes.
(Strictly speaking, in the smooth case it is necessary to show that
local solutions can be patched together. This may be done using the
domain of dependence result of Theorem~\ref{th:smooth} below.)  Also,
by the symmetry of the problem, we may restrict attention to a
neighbourhood of $\theta=0$ and $t=0$. When we write `the axis' below,
we will mean the axis at $\theta=0$, although all arguments apply to
the other axis at $\theta=\pi$ as well.

% % % % % % % % % % % % % % % % % % % % % % % % % % % % % % % % % % % % % % % %

\subsection{Parametrisations}
\label{sec:param}

The orbit metric $\id\sigma^2$ may be parametrised in several ways. If
we use the parametrisation
\begin{equation}\label{eq:g-PQ}
  \id s^2 = e^A\,(-\id t^2 + \id\theta^2)
          + R \bigl[ e^P (\id\phi + Q\id\chi)^2 + e^{-P} \id\chi^2 \bigr],
\end{equation}
the Einstein equations decouple as two evolution equations involving
$P$ and $Q$ and two constraint equations involving $P$, $Q$ and $A$
\cite{Chrusciel:U1xU1}.  Let $(t,\theta)\in M\subset\reals^2$ with the
Minkowski metric $\eta=-\id{t}^2+\id\theta^2$ and
$(P,Q)\in{N}\subset\reals^2$ with the hyperbolic metric
$h=\id{P}^2+e^{2P}\id{Q}^2$. Given the map
$\Psi\colon{M}\ni(t,\theta)\mapsto(P,Q)\in{N}$, we define a section
$\id\Psi$ of $T^*M\otimes\Psi^*(TN)$ by
\begin{equation}\label{eq:dPsi}
  \id\Psi = \id{P}\otimes\Psi^*\dd{P} + \id{Q}\otimes\Psi^*\dd{Q}.
\end{equation}
Then the evolution equations may be written as the wave-map type
equation
\begin{equation}\label{eq:wavemap}
  \tr_\eta \bigl( \D(h,\eta)(R\,\id\Psi) \bigr) = 0,
\end{equation}
where $\D(h,\eta)$ is a connection on $T^*M\otimes\Psi^*(TN)$ defined
in terms of the connections of $(M,\eta)$ and $(N,h)$ (see
\cite{Chrusciel:U1xU1} for details).

The `velocity' $\nu(t,\theta_0)$ is defined up to a sign as the
$h$-velocity of the solution $(P,Q)$ of \eqref{eq:wavemap} along a
curve $\theta=\theta_0$, rescaled by an appropriate factor.  In the
$\torus^3$ case, this factor is $t$, but in our case it is more
convenient to take the factor to be $\tan{t}$.  So we define the
velocity to be
\begin{equation}\label{eq:velocity-PQ}
  \nu = \sqrt{ (DP)^2 + e^{2P} (DQ)^2 },
\end{equation}
where $D=\tan{t}\;\d/\d{t}$.

An `asymptotically velocity dominated' (AVD) solution is a solution
which asymptotically approach a solution to the `AVD equations',
obtained by dropping pure spatial derivative terms in the evolution
equations
\cite{Eardley-Liang-Sachs:VTD-dust,Isenberg-Moncrief:Gowdy-polar-asymptotic}.
For an AVD solution, it follows immediately from the AVD equations
that the velocity $\nu$ tends to a finite limit as $t\to0$.  We will
refer to this limit as the asymptotic velocity.  It is also possible
to fix the sign of the velocity to be the same as the sign of $DP$,
note however that this sign depends upon the parametrisation as we
will see below.

Note that $P$ cannot be regular at the axis.  The regular
parametrisation used by Garfinkle in his numerical work
\cite{Garfinkle:Gowdy-S2xS1-num} is obtained from \eqref{eq:g-PQ} by
setting
\begin{equation}\label{eq:g-Garfinkle}
  P = \hat{P} + \ln\sin\theta
  \qquad\text{and}\qquad
  A = \hat{P} + \gamma + \ln\sin{t},
\end{equation}
with $\gamma$, $\hat{P}$ and $Q$ smooth functions of $t$ and $\theta$.
Unfortunately, the corresponding evolution equations contain
coefficients that are singular at the axes.  This is not a major
obstacle for the numerics since the full terms appearing in the
equations are well behaved, but it does prevent a direct analytical
treatment.

Another parametrisation may be obtained by interchanging the roles of
the Killing vectors $\d/\d{\phi}$ and $\d/\d{\chi}$.  This corresponds
to an inversion in the hyperbolic plane $(N,h)$, so the field
equations are invariant under this reparametrisation.  Explicitly,
\begin{equation}\label{eq:g-PQ-invert}
  \id s^2 = e^A\,(-\id t^2 + \id\theta^2)
          + R \bigl[ e^{Y} (\id\chi + X\id\phi)^2
                     + e^{-Y} \id\phi^2 \bigr],
\end{equation}
where
\begin{equation}\label{eq:inversion}
  Y = P + \ln(e^{-2P}+Q^2)
  \qquad\text{and}\qquad
  X = \frac{Q}{e^{-2P}+Q^2}.
\end{equation}
Note that there is a subtlety here. If $P\to\infty$ as $t\to0$, the
rate of blow-up of $Y$ will be dramatically different depending on
whether $Q$ vanishes or not. This is a consequence of the
parametrisation, and must be distinguished from real geometric
features of the model.

The parametrisation \eqref{eq:g-PQ-invert} is again singular at the
axis, but this may be resolved by substituting $Y=Z-\ln{R}$. We also
set $A=\lambda-Z+2\ln\sin{t}$. The metric then becomes
\begin{equation}\label{eq:g}
  \id s^2 = e^{\lambda-Z}\sin^2{t}\,(-\id t^2 + \id\theta^2) 
          + e^Z (\id\chi + X \id\phi)^2 + R^2 e^{-Z} \id\phi^2.
\end{equation}
From Lemma~5.1 of \cite{Chrusciel:U1xU1}, \eqref{eq:g} defines a
smooth metric with the desired symmetry if and only if $\lambda$, $Z$
and $X$ are smooth functions of $\theta$ on $\sphere^2$, with $X$ and
$\lambda$ vanishing at $\theta=0$ and $\theta=\pi$. Note that a
function is smooth in a neighbourhood of $\theta=0$ or $\theta=\pi$ in
$\sphere^2$ if and only if it is a smooth function of $\sin^2\theta$
there.

Because of the invariance of $h$ under inversion, the velocity is
\begin{equation}\label{eq:velocity}
  \nu = \sqrt{ (DY)^2 + e^{2Y} (DX)^2 }
      = \sqrt{ (1-DZ)^2 + e^{2Z} R^{-2} (DX)^2 }.
\end{equation}
As mentioned above, it is possible to assign a sign to $\nu$ by
setting it equal to the sign of $DP$. Assume, for illustrative
purposes, that $Q$ is independent of $t$ and that $DP\to\infty$ as
$t\to0$. Then it follows from \eqref{eq:inversion} that $DZ\to DP$ as
$t\to0$ wherever $Q\ne0$, but $DZ\to-DP$ for those values of $\theta$
where $Q$ vanishes. So the sign defined with respect to $DP$ cannot be
expressed in terms of $DZ$ alone. Also, if $DP$ tends to a smooth
function of $x$ as $t\to0$, the corresponding limit of $DZ$ has
discontinuities at points where $Q$ has isolated zeros (note that for
any smooth metric of the form \eqref{eq:g-PQ}, $Q$ must vanish at the
axes). Since $DP$ has a perfectly regular asymptotic behaviour in this
case, these discontinuities are artefacts of the parametrisation
(called `false spikes' in \cite{Rendall-Weaver:manufacture}). The same
argument may of course also be applied in the other direction, giving
discontinuities in the asymptotic behaviour of $DP$ from a
well-behaved $DZ$. For the rest of this paper, we fix the sign of the
velocity to be equal to the sign of $1-DZ$.

% % % % % % % % % % % % % % % % % % % % % % % % % % % % % % % % % % % % % % % %

\subsection{Einstein's equations}
\label{seq:equations}

In the last section we obtained a parametrisation which is regular at
the axis.  We will now modify the parametrisation to obtain evolution
equations with regular coefficients as well.

We will denote partial derivatives by subscripts, e.g.,
$f_\theta:=\d_{\theta}f=\d{f}/\d\theta$.  As mentioned above,
Einstein's equations decompose into the wave map equation
\eqref{eq:wavemap} for $Z$ and $X$ and two constraint equations for
$\lambda$, $Z$ and $X$.  If we denote the covariant derivative of the
Minkowski metric $\eta$ on $N$ by $\D$, \eqref{eq:wavemap} can be
written as
\begin{subequations}\label{eq:ee-Gowdy}
\begin{align}
\label{eq:ee1-Gowdy}
  R^{-1} \D_A(R\,\D^A Z) &= R^{-2}e^{2Z}\,\D_A X\,\D^A X,\\
\label{eq:ee2-Gowdy}
  R\,\D_A(R^{-1} e^{2Z} \D^A X) &= 0,
\end{align}
\end{subequations}
and the constraint equations are
\begin{equation}\label{eq:ec-Gowdy}
  2 (\lambda_\pm \pm \cot{t}) R_\pm
  = 4(R_\theta)_\pm + R (Z_\pm^2 + e^{2Z} R^{-2} X_\pm^2),
\end{equation}
where we have used the notation $f_\pm:=f_\theta\pm f_t$
\cite{Chrusciel:U1xU1}. 

The integrability of the constraints \eqref{eq:ec-Gowdy} are ensured
by the evolution equations \eqref{eq:ee-Gowdy}.  When $R_+$ and $R_-$
are nonzero, the constraints determine $\lambda$ up to a constant, and
the constant is fixed by the smoothness requirement, i.e., that
$\lambda=0$ at $\theta=0$. Since $R=\sin{t}\sin\theta$, $R_+$ and
$R_-$ vanish when $\theta=t$ and $\theta=\pi-t$ respectively. So this
happens for two values of $\theta$ in each Cauchy surface. At these
points, the constraints become `matching conditions' for $Z$ and $X$
\cite{Gowdy:Gowdy,Tanimoto:local-U1xU1}. If the matching conditions
hold on an initial Cauchy surface, they are preserved by the evolution
equations \cite{Garfinkle:Gowdy-S2xS1-num}. In what follows we will
assume without further comment that the solutions are chosen such that
the matching conditions hold.

To obtain a system with smooth coefficients, we reparametrise the
metric using an Ernst potential as in \cite{Chrusciel:U1xU1}, a
procedure which is similar to the Kramer-Neugebauer transformation in
stationary and axisymmetric spacetimes
\cite{Kramer-Neugebauer:axialsymm}.  Let
\begin{equation}\label{eq:Ernst}
  \Omega := -R^{-1} e^{2Z} (X_\theta\,\id t + X_t\,\id\theta).
\end{equation}
It then follows from \eqref{eq:ee2-Gowdy} that $\Omega$ is a closed
form.  Thus we may write $\Omega=\id\omega$ for some function
$\omega$, defined up to a constant, on any simply connected open set
containing $\theta=0$.  Inverting the relation \eqref{eq:Ernst} shows
that if $\omega$ is a smooth function of $\theta$ on a neighbourhood
of $\theta=0$ on $\sphere^2$, so is $X$.  This determines $X$ up to a
constant, which is fixed by the requirement that $X=0$ at $\theta=0$
so that the metric is smooth at the axis.  Following
\cite{Rendall-Weaver:manufacture}, we will call the transformation
$(Y,X)\mapsto(Z,\omega)$ a `Gowdy-to-Ernst' transformation.

Expressing the evolution equation \eqref{eq:ee1-Gowdy} in terms of $Z$
and $\omega$ gives
\begin{subequations}\label{eq:ee-Ernst}
\begin{equation}
  R^{-1} \D_A(R\,\D^A Z) = -e^{-2Z}\,\D_A\omega\,\D^A\omega,
\end{equation}
and the identity $\D_{[A}\D_{B]}X=0$ together with \eqref{eq:Ernst}
gives a second evolution equation
\begin{equation}
  \D_A(R e^{-2Z} \D^A \omega) = 0.
\end{equation}
\end{subequations}
Expanding \eqref{eq:ee-Ernst} we get 
\begin{subequations}\label{eq:ee-Ernst2}
\begin{align}
  Z_{tt} + \cot{t}\,Z_t - Z_{\theta\theta} - \cot\theta\,Z_\theta
  &= - e^{-2Z} (\omega_t^2 - \omega_\theta^2), \\
  \omega_{tt} + \cot{t}\,\omega_t
  - \omega_{\theta\theta} - \cot\theta\,\omega_\theta
  &= 2(Z_t\,\omega_t - Z_\theta\,\omega_\theta).
\end{align}
\end{subequations}

We will now rewrite the equations in a form which makes the regularity
of the coefficients at the axis explicit, and which is more suited for
the Fuchsian techniques.  Put $\tau:=\sin{t}$ and
$D:=\tan{t}\,\d_t=\tau\,\d_\tau$ as before, and let
$\lap:=\d_\theta^2+\cot\theta\,\d_\theta+(\sin\theta)^{-2}\d_\phi^2$,
e.g., the Laplacian on $\sphere^2$ with respect to the metric
$\id\theta^2+\sin^2\theta\,\id\phi^2$ induced by the natural embedding
in Euclidean $\reals^3$.  We also write
$\grad{f}\cdot\grad{g}:=f_\theta\,g_\theta+(\sin\theta)^{-2}f_\phi\,g_\phi$
and $(\grad{f})^2:=\grad{f}\cdot\grad{f}$.  Since $Z$ and $\omega$ are
independent of $\phi$, \eqref{eq:ee-Ernst2} can be written as
\begin{subequations}\label{eq:ee-final}
\begin{align}
  (1-\tau^2) D^2{Z} - \tau^2 DZ &= \tau^2 \lap{Z}
  - e^{-2Z}\bigl[(1-\tau^2)(D\omega)^2 - \tau^2(\grad\omega)^2\bigr], \\
  (1-\tau^2) D^2{\omega} - \tau^2 D\omega &= \tau^2 \lap\omega
  + 2\bigl[(1-\tau^2)D{Z}D\omega - \tau^2\,\grad{Z}\cdot\grad\omega\bigr].
\end{align}
\end{subequations}
The constraints \eqref{eq:ec-Gowdy} expressed in terms of $Z$ and
$\omega$ are
\begin{subequations}\label{eq:ec-Ernst}
\begin{align}
  \label{eq:ec-Ernst-1}
  D\lambda + \tan^2{t}\cot\theta\,\lambda_\theta &= -2(1+\tan^2{t}) 
  + \frac12 \Bigl[(DZ)^2 + \tan^2{t}\,Z_\theta^2 
    + e^{-2Z}\bigl((D\omega)^2 + \tan^2{t}\,\omega_\theta^2\bigr)\Bigr], \\
  \label{eq:ec-Ernst-2}
  D\lambda + \tan\theta\,\lambda_\theta &=  
    \tan\theta\,\bigl( Z_\theta\,DZ + e^{-2Z}\omega_\theta\,D\omega \bigr).
\end{align}
\end{subequations}

In the analytic case, the Fuchsian techniques can in fact be applied
directly to \eqref{eq:ee-final} in the variables $\tau$ and $\theta$.
For the smooth case however, we have to cast the equations into
symmetric hyperbolic form, which cannot be done using the $\theta$
variable alone because the axial symmetry will lead to coefficients
singular in $\theta$ at $\theta=0$.  We therefore introduce an
approximately Cartesian coordinate system on a neighbourhood of
$\theta=0$ on $\sphere^2$.  Let $x:=\sin\theta\cos\phi$ and
$y:=\sin\theta\sin\phi$.  In the variables $(x,y)$, the metric on
$\sphere^2$ is
\begin{equation}\label{eq:metric-S2-cart}
  \id\theta^2+\sin^2\theta\,\id\phi^2
  = (1-x^2-y^2)^{-1} 
    \bigl[ (1-y^2)\id{x}^2 + 2xy\,\id{x}\id{y} + (1-x^2)\id{y}^2 \bigr],
\end{equation}
hence
\begin{equation}\label{eq:Lap-S2-cart}
  \lap = (1-x^2)\,\d_x^2 - 2xy\,\d_x\d_y + (1-y^2)\,\d_y^2 
       - 2x\,\d_x - 2y\,\d_y,
\end{equation}
and
\begin{equation}\label{eq:grad2-S2-cart}
  \grad{f}\cdot\grad{g} 
  = (1-x^2){f_x}{g_x} - xy({f_x}{g_y}+{f_y}{g_x}) + (1-y^2){f_y}{g_y}.
\end{equation}
Inserting these expressions in \eqref{eq:ee-final} then gives a system
in the coordinates $t$, $x$ and $y$ with coefficients regular in $x$
and $y$ at the axis.

% % % % % % % % % % % % % % % % % % % % % % % % % % % % % % % % % % % % % % % %

\subsection{Restrictions at the axes}
\label{sec:axes}

As mentioned in the introduction, asymptotic velocity dominance for
Gowdy spacetimes may be loosely formulated as that the metric tends to
a Kasner metric when $t\to0$ for a fixed point in $\Sigma$. The
presence of symmetry axes imposes additional restrictions on the
asymptotic behaviour along the axis.

To make this precise we calculate the generalised Kasner exponents.
These are defined as the eigenvalues $q_i$ of the renormalised second
fundamental form $(\tr{k})^{-1}k_{ij}$, expressed in an orthonormal
frame on $\Sigma$.  From the expression for the metric \eqref{eq:g} we
find that
\begin{subequations}\label{eq:Kasner-exp}
\begin{align}
  q_1     &= (D\lambda-DZ+2)/(D\lambda-DZ+4), \\
  q_{2,3} &= (1\pm\nu)/(D\lambda-DZ+4),
\end{align}
\end{subequations}
where $\nu$ is the velocity as given in \eqref{eq:velocity}.

Because of the rotational symmetry, the only possible values of the
Kasner exponents at the axis are $(0,0,1)$ and $(2/3,2/3,-1/3)$ since
the two eigenvalues corresponding to eigenvectors tangent to $\Sigma$
must agree. From the smoothness requirements $D\lambda$ and $DX$
vanish at the axis, and using our sign convention we have $\nu=1-DZ$
there. It then follows immediately from \eqref{eq:Kasner-exp} that for
the $(0,0,1)$ case, $DZ\to2$ and $\nu\to-1$ as $t\to0$ along the axis,
while for the $(2/3,2/3,-1/3)$ case, $DZ\to-2$ and $\nu\to3$.

As noted at the beginning of this section, the $\torus^3$ results of
\cite{Kichen-Rendall:Gowdy-T3-analytic} applies to sets not containing
the axis at $\theta=0$.  In that case, we obtain solutions with the
full number of free functions only in the case when $0<\nu<1$ (and
$DP<0$) in the limit $t\to0$.  Also, numerical simulations have shown
that the velocity is indeed driven into this range, except at isolated
spatial points \cite{Berger-Moncrief:Gowdy-T3-num} (see also
\cite{Rendall-Weaver:manufacture} for a heuristic argument).  So we
cannot hope to show the existence of velocity dominated solutions
depending on the full number of free functions since $\nu$ is either
$-1$ or $3$ on the axis.  (Actually, in our case $DZ$ will play the
role of $\nu$, but the same conclusion can be drawn since $\nu=1-DZ$
on the axis.)

% % % % % % % % % % % % % % % % % % % % % % % % % % % % % % % % % % % % % % % %

\subsection{Some known solutions}
\label{sec:knownsol}

\subsubsection{Polarised solutions}
\label{sec:polarsol}

The polarised solutions form the subclass with $Q\equiv0$, so called
because they describe spacetimes with polarised gravitational waves.
In this case the evolution equations \eqref{eq:ee-final} reduce to a
linear ordinary differential equation for $P$, the
Euler-Poisson-Darboux equation.  The solutions may be expressed
explicitly in terms of Legendre polynomials
\cite{Hanquin-Demaret:Gowdy-S2xS1+S3}. Also, rigorous asymptotic
expansions in a neighbourhood of the singularity have been found
\cite{Isenberg-Moncrief:Gowdy-polar-asymptotic,%
Chrusciel-Isenberg-Moncrief:Gowdy-SCC}.
In our notation,
\begin{subequations}
\begin{align}
  Z(t,\theta) &= k(\theta)\ln\sin t + \varphi(\theta) + u(t,\theta), \\
  \lambda(t,\theta) &= \frac12(k^2-4)\ln\sin t + \psi(\theta) + v(t,\theta),
\end{align}
\end{subequations}
with $\abs{D^m\d_\theta^n{u}}\le{C}\sin^2{t}\abs{\ln\sin{t}}$ and
$\abs{D^m\d_\theta^n{v}}\le{C}\sin^2{t}\abs{\ln\sin{t}}^2$ for
$m=0,1$, all $n$ and a constant $C$.  The functions $k$ and $\varphi$
may be chosen freely as long as they satisfy the constraints, and the
solutions are all asymptotically velocity dominated. Since $\lambda$
must vanish at the axis for the metric to be smooth, it follows
immediately from the expansions that $k=\pm2$ there, and the velocity
$\nu=1-k$ must tend to $-1$ or $3$ along the axis. The Kretschmann
curvature scalar is unbounded as $t\to0$ unless $\nu=1-k=\pm1$,
$k_\theta=0$ and $k_{\theta\theta}=0$.

There is a way in which one may obtained unpolarised solutions from a
given polarised solution by means of an Ehlers transformation.  This
is used to obtain the `reference solution' used to validate the
numerical code in \cite{Garfinkle:Gowdy-S2xS1-num} (see also
\cite{Griffiths-Alekseev:unpolar-Gowdy}).  The technique works because
a combination of an Ehlers transformation with Gowdy-to-Ernst
transformations gives an isometry of the hyperbolic plane $(N,h)$.
The velocity of the transformed solution has the same asymptotic
behaviour as for the original polarised solution.

\subsubsection{Black hole solutions}
\label{sec:BHsol}

The region of Kerr spacetime between the inner and outer horizons may
be written as a Gowdy spacetime over $\sphere^2\times\sphere^1$, by
choosing as time coordinate a rescaling of the usual radial coordinate
\cite{Obregon-Ryan:Gowdy-S2xS1-Kerr}.  In our parametrisation
\eqref{eq:g}, the velocity $\nu$ tends to $-1$ along the axis and to
$+1$ elsewhere.  The discontinuous behaviour is an artefact of the
parametrisation, since in the parametrisation \eqref{eq:g-PQ},
$DP\to-1$ everywhere.  This provides an example of Gowdy spacetimes on
$\sphere^2\times\sphere^1$ without curvature singularities.  The
extremal case corresponding to the interior of the Schwarzschild black
hole is a polarised Gowdy solution with $\nu\to-1$ at the horizon and
$\nu\to3$ at the curvature singularity.

The same idea can be applied to the `distorted black hole' spacetimes
studied by Geroch and Hartle.  These are constructed by analytic
continuation of certain Weyl solutions, essentially corresponding to
perturbations on an exterior Schwarzschild background, through the
horizon.  The solutions are static, but they can be different from
Schwarzschild spacetime by violating asymptotic flatness or by
allowing matter in the exterior region.  The interior is a polarised
Gowdy spacetime with $\sphere^2\times\sphere^1$ topology. If we let
the Schwarzschild solution be given in Gowdy coordinates with $P=P_S$,
the Geroch-Hartle solution is given by $P=P_S+U$, where $U$ is regular
on the horizon. It is possible to show that $U$ must be regular at the
other singularity as well. Thus the asymptotic velocity is the same as
for the interior Schwarzschild solution.

Note that for polarised Gowdy spacetimes, analyticity is a necessary
as well as sufficient condition for the existence of an extension with
a compact Cauchy horizon \cite{Chrusciel-Isenberg-Moncrief:Gowdy-SCC}.
The necessity follows from the fact that on the extension, the
evolution equation becomes the Laplace equation in suitably chosen
coordinates.

\subsubsection{Numerical solutions}
\label{sec:numsol}

Garfinkle has done numerical simulations of Gowdy spacetimes on
$\sphere^2\times\sphere^1$, with similar results as in the $\torus^3$
case \cite{Garfinkle:Gowdy-S2xS1-num}.  All solutions have velocity
$-1$ at the axis, without spiky features, which probably can be
explained by the particular choice of initial data.  We should
emphasise however that Garfinkle's calculations are carried out in the
parametrisation \eqref{eq:g-Garfinkle}, and rewriting his results in
our parametrisation \eqref{eq:g} gives solutions with `false spikes'
at the axis.  Since we cannot generate solutions with spikes by the
method in this paper, our results cannot be compared directly with
those in \cite{Garfinkle:Gowdy-S2xS1-num}.

%%%%%%%%%%%%%%%%%%%%%%%%%%%%%%%%%%%%%%%%%%%%%%%%%%%%%%%%%%%%%%%%%%%%%%%%%%%%%%%

\section{The analytic case}
\label{sec:analytic}

% % % % % % % % % % % % % % % % % % % % % % % % % % % % % % % % % % % % % % % %

\subsection{Fuchsian systems}
\label{sec:fuchsian}

Consider a system of partial differential equations on $\reals^{n+1}$,
whose solutions are expected to have a singularity as $t\to0$. The
Fuchsian algorithm is based on the following idea: decompose the
unknown into a prescribed singular part, depending on a number of
arbitrary functions, and a regular part $w$. If the system can be
rewritten as a Fuchsian system of the form
\begin{equation}\label{eq:Fuchsian}
  t\d_t w + N(x)w = t f(t,x,w,w_x),
\end{equation}
where $w_x$ denotes the collection of spatial derivatives of $w$, the
Fuchsian method applies. In the analytic case we have the following
theorem.
\begin{theorem}%
  [Kichenassamy and Rendall \cite{Kichen-Rendall:Gowdy-T3-analytic}]
  \label{th:Fuchsian}
  Assume that $N$ is an analytic matrix near $x=0$ such that there is
  a constant $C$ with $\norm{\sigma^N}\le{C}$ for $0<\sigma<1$, where
  $\sigma^N$ is the matrix exponential of $N\ln\sigma$. Also, suppose
  that $f$ is a locally Lipschitz function of $w$ and $w_x$ which
  preserves analyticity in $x$ and continuity in $t$. Then the
  Fuchsian system \eqref{eq:Fuchsian} has a unique solution in a
  neighbourhood of $x=0$ and $t=0$ which is analytic in $x$ and
  continuous in $t$, and tends to $0$ as $t\to0$.
\end{theorem}

The method of proof is a variation of the standard Cauchy problem.
Note that Theorem~\ref{th:Fuchsian} also applies to the case
\begin{equation}\label{eq:Fuchsian-ta}
  t\d_t w + N(x)w = t^\alpha f(t,x,w,w_x),
\end{equation}
since changing the $t$ variable to $t^\alpha$ transforms the system
into the form \eqref{eq:Fuchsian}.

% % % % % % % % % % % % % % % % % % % % % % % % % % % % % % % % % % % % % % % %

\subsection{Low velocity, analytic case}
\label{sec:anal-low}

Let $k$, $\varphi$, $\omega_0$ and $\psi$ be real analytic functions
of $\theta$ on a neighbourhood of $\theta=0$ in $\sphere^2$, with
$k\in(0,1)$.  We introduce new unknowns $u$ and $v$ such that
\begin{subequations}\label{eq:expansion-low}
\begin{align}\label{eq:expansion-low-Z}
  Z(\tau,\theta) &= k(\theta)\ln\tau + \varphi(\theta) 
    + \tau^\epsilon u(\tau,\theta), \\
  \label{eq:expansion-low-omega}
  \omega(\tau,\theta) &= \omega_0(\theta) 
    + \tau^{2k(\theta)}\bigl(\psi(\theta) + v(\tau,\theta)\bigr),
\end{align}
\end{subequations}
where $\epsilon>0$ is a small constant which we will fix later.  The
idea is to use Theorem~\ref{th:Fuchsian} to show that $u$, $v$, $Du$
and $Dv$ tend to $0$ as $t\to0$, so that \eqref{eq:expansion-low}
provides an asymptotic expansion of $Z$ and $\omega$.  The term `low
velocity' is then justified by the fact that the velocity $\nu$ tends
to $1-k\in(0,1)$ as $t\to0$, as is easily verified from the definition
\eqref{eq:velocity}.  For this reason we will refer to $1-k$ as the
`asymptotic velocity'.  However, remember that from the discussion in
section~\ref{sec:axes}, the velocity should be $-1$ or $3$ at the
axis, so the solutions constructed in this section are \emph{not}
regular at the axis.

In terms of $X$, the expansion \eqref{eq:expansion-low-omega} of $\omega$ 
corresponds to
\begin{equation}\label{eq:expansion-low-X}
  X = X_0(\theta) 
    + \tau^{2(1-k(\theta))}\bigl(\tilde{\psi}(\theta) + 
                                 \tilde{v}(\tau,\theta)\bigr),
\end{equation}
for some functions $X_0$, $\tilde{\psi}$ and $\tilde{v}$, which can be
compared directly with the $\torus^3$ results
\cite{Kichen-Rendall:Gowdy-T3-analytic}.

Inserting the expansion \eqref{eq:expansion-low} into the system
\eqref{eq:ee-final} gives
\begin{subequations}\label{eq:ee-exp-low}
\begin{align}
\begin{split}
  (1-\tau^2) D^2 u 
  =& -(1-\tau^2) \bigl[ 2\epsilon Du + \epsilon^2 u \bigr]
  + \tau^{2-\epsilon} \bigl[ k + (\ln\tau)\lap{k} + \lap\varphi \bigr]
  + \tau^2 \ignorewidth{ \bigl[ (D+\epsilon)u + \lap{u} \bigr] } \\ % ignorew.
  &- e^{-2\varphi-2\tau^{\epsilon}u} 
      \Bigl\{
        (1-\tau^2)\tau^{2k-\epsilon} \bigl[(D+2k)(\psi+v)\bigr]^2
        - \tau^{2-2k-\epsilon}(\grad\omega_0)^2 \\
  &\phantom{- e^{-2\varphi-2\tau^{\epsilon}u} \Bigl\{}
        - 2\tau^{2-\epsilon} \grad\omega_0 \cdot
            \bigl[ \grad(\psi+v) + 2(\ln\tau)(\psi+v)\grad{k} \bigr] \\
  &\phantom{- e^{-2\varphi-2\tau^{\epsilon}u} \Bigl\{}
      - \tau^{2+2k-\epsilon} 
          \bigl[ \grad(\psi+v) + 2(\ln\tau)(\psi+v)\grad{k} \bigr]^2
      \Bigr\},
\end{split} \displaybreak[0]\\
\begin{split}
  (1-\tau^2) D^2 v =& -2k (1-\tau^2) Dv 
  + 2(1-\tau^2)\tau^\epsilon(D+\epsilon)u(D+2k)(\psi+v) \\
  &+ \tau^2 \bigl[
              2(\ln\tau)\grad{k}\cdot\grad(\psi+v)
              + 2(\ln\tau)(\psi+v)\lap{k} \\
  &\phantom{+ \tau^2 \Bigl\{}
              + (D+2k)(\psi+v)
              + \lap(\psi+v)
            \bigr] \\
  &+ \tau^{2-2k}\lap\omega_0
  - 2\tau^{2-2k}\grad\omega_0\cdot
      \bigl[ 
        (\ln\tau)\grad{k} + \grad\varphi + \tau^\epsilon \grad{u}
      \bigr] \\
  &- 2\tau^2 
      \bigl[ \grad\varphi + \tau^\epsilon \grad{u} \bigr]
      \cdot
      \bigl[ 2(\ln\tau)(\psi+v)\grad{k} + \grad(\psi+v) \bigr].
\end{split}
\end{align}
\end{subequations}

Next, we introduce the variables
\begin{subequations}\label{eq:UVdef}
\begin{alignat}{4}
  U_0 &:= u, &\qquad 
  U_1 &:= Du, &\qquad
  U_2 &:= \tau u_x, &\qquad
  U_3 &:= \tau u_y, \\
  V_0 &:= v, &\qquad
  V_1 &:= Dv, &\qquad
  V_2 &:= \tau v_x, &\qquad 
  V_3 &:= \tau v_y.
\end{alignat}
\end{subequations}
We also write $\Ubar$ and $\Vbar$ as shorthands for the vectors
$(U_2,U_3)$ and $(V_2,V_3)$. Using \eqref{eq:Lap-S2-cart} and
\eqref{eq:grad2-S2-cart} the system \eqref{eq:ee-exp-low} may then be
written as
\begin{subequations}\label{eq:Fuchsian-low}
\begin{align}
  DU_0 =&\; U_1, \\
  \begin{split}\label{eq:Fuchsian-low-DU1}
    (1-\tau^2) DU_1 
    =& -(1-\tau^2) \bigl[ 2\epsilon U_1 + \epsilon^2 U_0 \bigr]
    + \tau^{2-\epsilon} \bigl[ k + (\ln\tau)\lap{k} + \lap\varphi \bigr]
    + \ignorewidth{ \tau^2 (U_1 + \epsilon U_0) } \\% ignorewidth
    &+ \tau \bigl[
              (1-x^2)U_{2x} - xy(U_{2y}+U_{3x}) + (1-y^2) U_{3y}
              - 2x U_2 - 2y U_3
            \bigr] \\
  &- e^{-2\varphi-2\tau^{\epsilon}U_0} 
      \Bigl\{
        (1-\tau^2)\tau^{2k-\epsilon} \bigl[V_1 + 2k(\psi+V_0)\bigr]^2
        - \tau^{2-2k-\epsilon}(\grad\omega_0)^2 \\
  &\phantom{- e^{-2\varphi-2\tau^{\epsilon}U_0} \Bigl\{}
        - 2\tau^{1-\epsilon} \grad\omega_0 \cdot
            \bigl[
              \Vbar + \tau\grad\psi + 2\tau(\ln\tau)(\psi+V_0)\grad{k}
            \bigr] \\
  &\phantom{- e^{-2\varphi-2\tau^{\epsilon}U_0} \Bigl\{}
        - \tau^{2k-\epsilon} 
            \bigl[
              \Vbar + \tau\grad\psi + 2\tau(\ln\tau)(\psi+V_0)\grad{k}
            \bigr]^2
      \Bigr\},
  \end{split} \\
  DU_2 =&\; \tau(U_{0x}+U_{1x}), \\
  DU_3 =&\; \tau(U_{0y}+U_{1y}), \displaybreak[0]\\
  DV_0 =&\; V_1, \\
  \begin{split}\label{eq:Fuchsian-low-DV1}
    (1-\tau^2) DV_1 =& -2k (1-\tau^2) V_1 
    + 2(1-\tau^2)\tau^\epsilon
        (U_1+\epsilon U_0) \bigl( V_1+2k(\psi+V_0) \bigr) \\
    &+ \tau^2 \ignorewidth{ \bigl[% begin ignorewidth
                V_1 + 2k(\psi+V_0)
                + 2(\ln\tau)(\psi+V_0)\lap{k}
                + 2(\ln\tau)\grad{k}\cdot\grad\psi + \lap\psi
              \bigr] } \\% end ignorewidth
    &+ 2\tau(\ln\tau) \grad{k}\cdot\Vbar + \tau^{2-2k}\lap\omega_0 \\
    &+ \tau \bigl[
              (1-x^2)V_{2x} - xy(V_{2y}+V_{3x}) + (1-y^2) V_{3y}
              - 2x V_2 - 2y V_3
            \bigr] \\
    &- 2\tau^{2-2k}\grad\omega_0\cdot
        \bigl[ 
          (\ln\tau)\grad{k} + \grad\varphi + \tau^\epsilon \grad{U_0}
        \bigr] \\
    &- 2\tau 
        \bigl[ \tau\grad\varphi + \tau^\epsilon \Ubar \bigr]
        \cdot 
        \bigl[ 2(\ln\tau)(\psi+V_0)\grad{k} + \grad(\psi+V_0) \bigr],
  \end{split} \\
  DV_2 =&\; \tau(V_{0x}+V_{1x}), \\
  DV_3 =&\; \tau(V_{0y}+V_{1y}).
\end{align}
\end{subequations}

If we choose $\epsilon$ such that $0<\epsilon<\min\{2k,2-2k\}$, the
system \eqref{eq:Fuchsian-low} is of the form \eqref{eq:Fuchsian-ta}
for some $\alpha$ (after dividing \eqref{eq:Fuchsian-low-DU1} and
\eqref{eq:Fuchsian-low-DV1} by $1-\tau^2$), and a rescaling of $\tau$
gives a Fuchsian system of the form \eqref{eq:Fuchsian}.  By the
criterion in \cite{Andersson-Rendall:quiescent-sings}, or by direct
computation of $\sigma^N$, the conditions of Theorem~\ref{th:Fuchsian}
are fulfilled.  (Note that the essential features of the system, in
particular the matrix $N$ and the powers of $\tau$, are the same as in
the $\torus^3$ case \cite{Kichen-Rendall:Gowdy-T3-analytic}.)  We come
to the following conclusion.
\begin{theorem}\label{th:anal-low}
  Suppose that $k$, $\varphi$, $\omega_0$ and $\psi$ are real analytic
  functions of $\theta$ in a neighbourhood of $\theta=0$ in
  $\sphere^2$ such that $k\in(0,1)$ for each $\theta$. Let $\epsilon$
  be a positive constant less than $\min\{2k,2-2k\}$. There exists a
  unique solution of Einstein's equations \eqref{eq:ee-Gowdy} of the
  form \eqref{eq:expansion-low} in a neighbourhood of $\theta=0$ and
  $t=0$ such that $u$, $v$, $Du$ and $Dv$ all tend to $0$ as $t\to0$.
\end{theorem}

% % % % % % % % % % % % % % % % % % % % % % % % % % % % % % % % % % % % % % % %

\subsection{Negative velocity, analytic case}
\label{sec:anal-neg}

In the case when we only assume $k$ to be positive, allowing values of
$k$ greater than $1$, we may use the same expansion
\eqref{eq:expansion-low} as in the low velocity case.  However,
examining the system \eqref{eq:Fuchsian-low} we see that we must
assume that $\omega_0$ is constant in order to avoid negative powers
of $\tau$.  This is again similar to the $\torus^3$ case, and is to be
expected since numerical simulations indicate that spiky features may
appear as $t\to0$
\cite{Kichen-Rendall:Gowdy-T3-analytic,Garfinkle:Gowdy-S2xS1-num}.
Again, using the definition \eqref{eq:velocity} of the velocity shows
that the asymptotic velocity is $1-k$, which is negative if $k>1$.
The expansion \eqref{eq:expansion-low-omega} of $\omega$ implies
\begin{equation}
  X = X_0(\theta) + \tau^{\epsilon}\tilde{v}(\tau,\theta).
\end{equation}
for some $X_0$ and $\tilde{v}$, which is in agreement with
\cite{Kichen-Rendall:Gowdy-T3-analytic}.

By the same arguments as in the low velocity case we have a similar
existence theorem for a smaller set of data, including the regular
case with velocity $-1$ at the axis.
\begin{theorem}\label{th:anal-high}
  Suppose that $k$, $\varphi$ and $\psi$ are real analytic functions
  of $\theta$ in a neighbourhood of $\theta=0$ in $\sphere^2$ such
  that $k>0$ for each $\theta$, and suppose that $\omega_0$ is a
  constant. Let $\epsilon$ be a positive constant less than $2k$.
  There exists a unique solution of Einstein's equations
  \eqref{eq:ee-Gowdy} of the form \eqref{eq:expansion-low} in a
  neighbourhood of $\theta=0$ and $t=0$ such that $u$, $v$, $Du$ and
  $Dv$ all tend to $0$ as $t\to0$.
\end{theorem}

% % % % % % % % % % % % % % % % % % % % % % % % % % % % % % % % % % % % % % % %

\subsection{High velocity, analytic case}
\label{sec:anal-high}

If $k$ is negative, we replace the expansion \eqref{eq:expansion-low}
with
\begin{subequations}\label{eq:expansion-neg}
\begin{align}
  Z(\tau,\theta) &= k(\theta)\ln\tau + \varphi(\theta)
    + \tau^\epsilon u(\tau,\theta), \\
  \omega(\tau,\theta) &= \omega_0(\theta) + \tau^\epsilon v(\tau,\theta).
\end{align}
\end{subequations}
Note that we do not get the full number of free functions in this case
either.  Again, this is because of the possibility of spiky features
in the asymptotic behaviour.  The calculations are similar to the
$k>0$ case and will not be repeated here.  It turns out that the
system is valid for $k<1/2$ (corresponding to an asymptotic velocity
greater than $1/2$), by choosing
$\max\{0,2k\}<\epsilon<\min\{2,2-2k\}$.  The corresponding expansion
of $X$ is
\begin{equation*}
  X = X_0 
    + \tau^{2(1-k(\theta))}(\tilde{\psi}(\theta)+\tilde{v}(\tau,\theta)),
\end{equation*}
for some functions $\tilde{\psi}$ and $\tilde{v}$ and an integration
constant $X_0$ (which must vanish for the solution to be smooth at the
axis). These solutions include the regular case with velocity $3$ at
the axis.
\begin{theorem}\label{th:anal-neg}
  Suppose that $k$, $\varphi$ and $\omega_0$ are real analytic
  functions of $\theta$ in a neighbourhood of $\theta=0$ in
  $\sphere^2$, such that $k<1/2$ for each $\theta$. Let $\epsilon$ be
  a positive constant such that
  $\max\{0,2k\}<\epsilon<\min\{2,2-2k\}$. There exists a unique
  solution of Einstein's equations \eqref{eq:ee-Gowdy} of the form
  \eqref{eq:expansion-neg} in a neighbourhood of $\theta=0$ and $t=0$
  such that $u$, $v$, $Du$ and $Dv$ all tend to $0$ as $t\to0$.
\end{theorem}

%%%%%%%%%%%%%%%%%%%%%%%%%%%%%%%%%%%%%%%%%%%%%%%%%%%%%%%%%%%%%%%%%%%%%%%%%%%%%%%

\section{The smooth case}
\label{sec:smooth}

In the previous section we established existence of real analytic
solutions with the desired asymptotic behaviour in a number of cases.
Here we will consider the corresponding smooth solutions, using a
generalisation of the approximation scheme in
\cite{Rendall:Gowdy-T3-smooth}.

The basic idea is to approximate smooth asymptotic data
$(k,\varphi,\omega_0,\psi)$ by a sequence of analytic data
$(k_m,\varphi_m,\omega_{0m},\psi_m)$, and show convergence of the
corresponding analytic solutions $w_m$ to a smooth solution $w$.
Since the argument does not depend on the details of the Gowdy models,
we will study a more general symmetric hyperbolic system with a
singular term and sufficiently well-behaved coefficients. Also, since
we are only interested in a neighbourhood of one of the axes in
$\sphere^2\times\sphere^1$ or $\sphere^3$, we will carry out the
argument on a subset of $[0,\infty)\times\reals^n$.
\begin{definition}\label{def:regsymmhyp}
  We say that the system of differential equations
  \begin{equation}\label{eq:symmhyp}
    t A^0(t,x)\d_t{w} + N(x)w + t A^j(t,x,w)\d_j{w} = t f(t,x,w)
  \end{equation}
  is \emph{regular symmetric hyperbolic} if $A^0$ is uniformly
  positive definite and symmetric, the $A^j$ are symmetric, and all
  coefficients are assumed to be regular in a sense to be specified
  below.
\end{definition}
\noindent
Note that our terminology differs from the usual one here, since the
full system \eqref{eq:symmhyp} is not regular at $t=0$ in the usual
sense.  In \cite{Rendall:Gowdy-T3-smooth}, the case when $A^0$ is the
identity was considered. We will therefore only give an outline of how
to generalise the argument to the case of more general $A^0$.

In our case, the system \eqref{eq:symmhyp} can be written as the
Fuchsian system
\begin{equation}\label{eq:Fuchsian2}
  t\d_t w + \tilde{N}(x)w = t \tilde{f}(t,x,w,w_x).
\end{equation}
Indeed, if $A^0$ were independent of $t$, we could put
$\tilde{N}:=(A^0)^{-1}N$ and $\tilde{f}:=(A^0)^{-1}(f-A^j\d_j{w})$.
Now the $t$-dependence of $A^0$ in our case is such that we may write
$(A^0)^{-1}(t,x)=B_0(x)+t^{\alpha}B_1(t,x)$ for some regular matrices
$B_0$ and $B_1$ and a constant $\alpha$ for small enough $t$.  The $t$
dependence of $(A^0)^{-1}N$ may then be included in $\tilde{f}$ (after
rescaling $t$ if $\alpha<1$).  The techniques of
\cite{Rendall:Gowdy-T3-smooth} can then be applied to
\eqref{eq:Fuchsian2} to obtain formal solutions (see below).

Note also that most of the arguments below hold in the more general
case when $A^0$ depends on $w$ as well.  However, in that case it is
often not possible to rewrite the symmetric hyperbolic system as a
Fuchsian system of the form \eqref{eq:Fuchsian2}.

% % % % % % % % % % % % % % % % % % % % % % % % % % % % % % % % % % % % % % % %

\subsection{Regularity and formal solutions}
\label{sec:regular}

Here we recall some results from \cite{Rendall:Gowdy-T3-smooth} on the
existence of formal solutions.  We first need appropriate notions of
regularity and of formal solutions.
\begin{definition}\label{def:regular}
  A function $f(t,x)$ from an open subset
  $\Omega\subset[0,\infty)\times\reals^n$ to $\reals^m$ is called
  \emph{regular} if it is $C^\infty$ for all $t>0$ and if its partial
  derivatives of any order with respect to $x\in\reals^n$ extend
  continuously to $t=0$ (within $\closure{\Omega}$).
\end{definition}
\begin{definition}\label{def:formal}
  A finite sequence $(w_1,w_2,\dots,w_p)$ of functions defined on an
  open subset $\Omega\subset[0,\infty)\times\reals^n$ is called a
  \emph{formal solution of order $p$} of the differential equation
  \eqref{eq:Fuchsian2} on $\Omega$ if
  \begin{enumerate}
  \item each $w_i$ is regular and
  \item $t\d_t w_i + \tilde{N}(x)w_i - t \tilde{f}(t,x,w_i,w_{ix})
         = \Ordo(t^i)$ for all $i$ as $t\to0$ in $\Omega$.
  \end{enumerate}
\end{definition}

The existence of formal solutions to a system of the form
\eqref{eq:Fuchsian2} is provided by the following result.
\begin{lemma}[Rendall \cite{Rendall:Gowdy-T3-smooth}]\label{la:formal-sol}
  If $\tilde{f}$ is regular and $\tilde{N}$ is smooth and satisfies
  $\norm{\sigma^{\tilde{N}}}\le{C}$ for some constant $C$ and all
  $\sigma$ in a neighbourhood of $0$, then \eqref{eq:Fuchsian2} has a
  formal solution of any given order which vanishes at $t=0$.
\end{lemma}
\noindent
By the discussion in the previous section, a formal solution of
\eqref{eq:Fuchsian2} may also be regarded as a formal solution of
\eqref{eq:symmhyp} with a definition similar to
Definition~\ref{def:formal}. We have thus established the existence
of formal solutions to \eqref{eq:symmhyp} of any order.

In proving the existence of smooth solutions to \eqref{eq:symmhyp} it
is important that the matrix $N$ is positive definite.  This is not
the case in general, so we need to modify the system to fulfil this
requirement.  The idea is to use a formal solution $\{w_1,\dots,w_i\}$
and to study the system satisfied by $z_i:=t^{1-i}(w-w_i)$, where $w$
is the sought solution of the original system \eqref{eq:symmhyp}.  The
procedure is similar to that in \cite{Rendall:Gowdy-T3-smooth}, so the
details will be omitted.  We end up with a system
\begin{equation}\label{eq:symmhyp-diff}
  t A^0(t,x)\d_t{z_i} + \bigl(N(x)+(i-1)A^0(t,x)\bigr)z_i 
  + t A^j(t,x,w_i+t^{i-1}z_i)\d_j{z_i} = t f_i(t,x,z_i),
\end{equation}
where the regular function $f_i$ is constructed from $A^j$ and $f$ and
depend on $w_i$ and $w_{ix}$ as well. The point is that since $A^0$ is
uniformly positive definite, the coefficient $N+(i-1)A^0$ is positive
definite for large enough $i$. We will refer to
\eqref{eq:symmhyp-diff} as the positive definite system.

% % % % % % % % % % % % % % % % % % % % % % % % % % % % % % % % % % % % % % % %

\subsection{The existence theorem}
\label{sec:smooth-exist}

Given smooth asymptotic data $S:=\{k,\varphi,\omega_0,\psi\}$ on
$\U\subset\reals^n$, we may construct a sequence of analytic data
$S_m:=\{k_m,\varphi_m,\omega_{0m},\psi_m\}$ on $\U$ which converges to
$S$ in $C^\infty(\U)$, uniformly on compact subsets.  If the formal
solutions are constructed as in the proof of Lemma~\ref{la:formal-sol}
(see \cite{Rendall:Gowdy-T3-smooth}), the analytic formal solutions
$w_{mi}$ of order $i$ corresponding to the analytic data $S_m$
converge to a formal solution $w_i$ of order $i$ corresponding to the
smooth data $S$ as $m\to\infty$, and the convergence is uniform on
compact subsets. This also holds for spatial derivatives of any order.
Hence spatial derivatives of any order of the coefficients of the
positive definite system \eqref{eq:symmhyp-diff} converge on compact
subsets as $m\to\infty$.  It follows that on any compact subset there
is an $i$ such that the coefficient involving $N(x)$ is positive
definite for all $m$ in our case, so we fix such a value of $i$ and
omit the index $i$ from now on.

The global existence theorem for Gowdy spacetimes over
$\sphere^2\times\sphere^1$ and $\sphere^3$ \cite{Chrusciel:U1xU1}
implies that there are smooth solutions of \eqref{eq:symmhyp-diff} on
a common time interval for all $m$. Thus our problem can be solved by
proving the following theorem.
\begin{theorem}\label{th:smooth}
  Let $z_m(t,x)$ be a sequence of regular solutions on
  $[0,t_1)\times\U\subset[0,\infty)\times\reals^n$, with
  $z_m(0,x)=0$, to a sequence of regular symmetric hyperbolic
  equations
  \begin{equation}\label{eq:symmhyp-diff2}
    t A_m^0(t,x)\d_t{z_m} + N_m(x)z_m + t A_m^j(t,x,z_m)\d_j{z_m}
    = t f_m(t,x,z_m).
  \end{equation}
  Suppose that $N_m$ is positive definite for each $m$ and that the
  coefficients converge uniformly to $A_0^0$, $N_0$, $A_0^j$ and $f_0$
  on compact subsets as $m\to0$, with the same properties as $A_m$,
  $N_m$, $A_m^j$ and $f_m$, and that the corresponding spatial
  derivatives converge uniformly as well. Then $z_m$ converges to a
  regular solution $z_0$ of the corresponding system with coefficients
  $A_0^0$, $N_0$, $A_0^j$ and $f_0$ on $[0,t_0)\times\U$ for some
  $t_0$, and $z_0(0,x)=0$.
\end{theorem}

\begin{proof}
  The idea is to use energy estimates to show that $\{z_m\}$ is a
  Cauchy sequence.  Since the proof is very similar to the case when
  $A^0$ is the identity \cite{Rendall:Gowdy-T3-smooth}, we only give
  an outline of the important steps here, with emphasis on the
  differences.
  
  First we consider the system satisfied by spatial derivatives of
  $z$. The collection $\z$ of spatial derivatives up to order $s$
  satisfies a system similar to \eqref{eq:symmhyp-diff2},
  \begin{equation}\label{eq:symmhyp-diff-s}
    t \A_m^0(t,x)\d_t{\z_m} + \N_m(x)\z_m
    + t \A_m^j(t,x,z_m)\d_j{\z_m} = t\,\f_m(t,x,z_m),
  \end{equation}
  obtained by differentiating \eqref{eq:symmhyp-diff2} and
  substituting the equations for lower order spatial derivatives.  The
  problem is that $\N_m$ is not necessarily positive definite, due to
  the presence of off-diagonal blocks depending on $(A_m^0)^{-1}$ and
  spatial derivatives of $A_m^0$ and $N_m$.  But this can be dealt
  with by multiplying the spatial derivatives $D^{\alpha}z_m$ by
  $K^{\abs\alpha}$ for a sufficiently small constant $K$ as in
  \cite{Rendall:Gowdy-T3-smooth} (here $\alpha$ is a multi-index).
  Let $\z_m:=\{K^{\abs\alpha}D^{\alpha}z_m;\abs\alpha\le{s}\}$ be the
  collection of weighted spatial derivatives up to order $s$ and
  assume that $\A_m$, $\N_m$, $\A_m^j$ and $\f_m$ are the
  corresponding weighted coefficient matrices.  For example, in one
  spatial dimension the system for $\z_m=(z_m,\d_x{z_m})$ has
  coefficients
  \begin{equation}
    \setlength{\arraycolsep}{3pt}
    \A_m^0 = 
    \begin{bmatrix}
      A_m^0 & 0  \\
      0     & A_m^0
    \end{bmatrix}\!\!,
    \ 
    \N_m =
    \begin{bmatrix}
      N_m       & 0  \\
      K D_m N_m & N_m
    \end{bmatrix}\!\!,
    \ 
    \A_m^j =
    \begin{bmatrix}
      A_m^j       & 0  \\
      K D_m A_m^j & A_m^j
    \end{bmatrix}\!\!,
    \ 
    \f_m =
    \begin{bmatrix}
      f_m  \\
      K D_m f_m
    \end{bmatrix}\!\!,
  \end{equation}
  where $D_m$ is the operator $\d_x - (\d_x A_m^0)(A_m^0)^{-1}$.  The
  convergence of the coefficients for any fixed $s$ follows from the
  convergence of the coefficients of the original equation.
  
  We will need appropriately weighted Sobolev norms
  \begin{equation}
    \normHs{u} :=
    (\sum_{\abs\alpha\le{s}} K^{2\abs\alpha}
    \iprodLii{D^{\alpha}{u}}{A_m^0{D^{\alpha}{v}}})^{1/2}.
  \end{equation}
  Note that these are equivalent to the usual Sobolev norms.  The
  norms depend a priori on $m$ since they include a factor $A_m^0$,
  but since $A_m^0$ converges uniformly on compact subsets to $A^0$
  which is uniformly positive definite, the equivalence can be taken
  to be independent of $m$ on compact subsets for sufficiently large
  $m$.
  
  Second, a domain of dependence result may be obtained by standard
  techniques.  The argument is the same as that in
  \cite{Rendall:Gowdy-T3-smooth} and will be omitted.  Thus we need
  only consider the problem on a compact subset.
  
  Next, we will show that the sequence $\normHs{z_m}$ is bounded.
  Differentiating and using the definitions of $\z_m$ and $\A_m^0$ we
  get
  \begin{equation}
    \d_t(\normHs{z_m}^2) 
    = 2 \iprodLii{\A_m^0\d_t\z_m}{\z_m}
    + \sum_{\abs\alpha\le s} K^{2\abs\alpha}
        \iprodLii{D^\alpha z_m}{(\d_t A_m^0)D^\alpha z_m},
  \end{equation}
  and using \eqref{eq:symmhyp-diff-s} to substitute for $\A_m^0\d_t\z_m$ 
  gives
  \begin{equation}\label{eq:zmBound}
    \d_t(\normHs{z_m}^2) = -2t^{-1}\iprodLii{\N_m\z_m}{\z_m} + R_m,
  \end{equation}
  where $R_m$ contains the same terms as in the regular case with
  $\N_m\equiv0$.  The first term on the right is negative since $\N_m$
  is positive definite by construction, and $R_m$ may be estimated as
  in the regular case, giving
  \begin{equation}\label{eq:RmsBound}
    R_m \le C_1 \normHs{z_m}^2 + C_2 \normHs{z_m}.
  \end{equation}
  Here $C_1$ and $C_2$ are polynomials in $\normLinf{(A^0)^{-1}}$,
  $\normLinf{D^{\alpha}A^0}$, $\normLinf{D^{\alpha}A^j}$ and
  $\normLinf{D^{\alpha}f}$ for $\abs\alpha\le{s}$, and in
  $\norm{z_m}_{H^k}$ for any given $k>n/2+1$ (by the Sobolev embedding
  theorem). Since all of the coefficients converge and we may choose
  $k=s$ if $s>n/2+1$, $\d_t(\normHs{z_m}^2)$ is bounded by a
  polynomial in $\normHs{z_m}$ whose coefficients are independent of
  $m$.  Applying Gronwalls lemma then gives that $\normHs{z_m}$ is
  bounded.
  
  We can in fact obtain a stronger estimate. Since $\normHs{z_m}$ is
  bounded, it follows that the coefficients $C_1$ and $C_2$ are
  bounded, so they can be chosen to be constants. A second application
  of Gronwalls lemma then gives that $t^{-1}\normHs{z_m}$ is bounded
  if $s>n/2+1$.
  
  Finally, we show that $\{z_m\}$ is a Cauchy sequence in the $H^s$
  norm. Let the difference between consecutive elements of the
  sequence be given by $v_m:=z_m-z_{m-1}$ and put
  $\vv_m:=\z_m-\z_{m-1}$.  From \eqref{eq:symmhyp-diff-s} we get
  \begin{equation}\label{eq:vmBound}
    \begin{split}
      \tfrac12\,\d_t(\normHs{v_m}^2) 
      =& - t^{-1}\iprodLii{\N_m\vv_m}{\vv_m} 
         - t^{-1}\iprodLii{(\N_m-\N_{m-1})\z_{m-1}}{\vv_m} \\
       & - \iprodLii{(\A_m^0-\A_{m-1}^0)\d_t\z_{m-1}}{\vv_m}
         - \iprodLii{\A_m^j(z_m)\d_j{\vv_m}}{\vv_m} \\
       & + \iprodLii{\vv_m}{\g_m},
    \end{split}
  \end{equation}
  where
  \begin{equation}\label{eq:gm}
    \g_m := t\bigl(\f_m(z_m)-\f_{m-1}(z_{m-1})\bigr)
          - t\bigl(\A_m^j(z_m)-\A_{m-1}^j(z_{m-1})\bigr)\d_j\z_{m-1}.
  \end{equation}
  The first term on the right hand side of \eqref{eq:vmBound} is
  negative and can be discarded, and using that
  $t^{-1}\normHs{z_{m-1}}$ is bounded, the second term is less than
  $C\normLinf{\N_{m+1}-\N_m}\normHs{v_m}$.  To estimate the third
  term we need a bound on $\normHs{\d_t{z_{m-1}}}$, but this can be
  obtained by applying the bound on $t^{-1}\norm{z_{m-1}}_{H^{s+1}}$
  to \eqref{eq:symmhyp-diff-s}.  The fourth term may be estimated by
  doing a partial integration as in the regular case.  The last term
  is estimated by inserting some terms and applying the mean value
  theorem and the bound on $\norm{z_{m-1}}_{H^{s+1}}$.  We end up
  with an estimate of the form
  \begin{equation}\label{eq:vmBound2}
    \d_t(\normHs{v_m}^2)
    \le C_1\normHs{v_m}^2+C_2\normHs{v_m},
  \end{equation}
  where $C_1$ is a constant independent of $m$ and $C_2$ is a
  polynomial in $\normLinf{\A_m^0-\A_{m-1}^0}$,
  $\normLinf{\N_m-\N_{m-1}}$, $\normLinf{\A_m^j-\A_{m-1}^j}$ and
  $\normLinf{\f_m-\f_{m-1}}$ which tends to $0$ as $m\to\infty$.
  Applying Gronwalls Lemma again, we conclude that $\normHs{v_m}\to0$
  as $m\to\infty$.
  
  We have shown that the solutions $z_m$ of the positive definite system
  \eqref{eq:symmhyp-diff2} converges in the $H^s$ norm to a smooth
  solution for each $s$.  A potential problem is that the time
  interval of convergence may depend on $s$.  But by standard
  techniques for symmetric hyperbolic equations the solutions may be
  extended to a common time interval, so we can conclude that $z_m$
  converges as smooth functions as well.
\end{proof}

% % % % % % % % % % % % % % % % % % % % % % % % % % % % % % % % % % % % % % % %

\subsection{Low velocity, smooth case}
\label{sec:smooth-low}

From the discussions in the previous section it follows that we only
need to write the Gowdy equations \eqref{eq:Fuchsian-low} in symmetric
hyperbolic form, since the existence of a smooth solution is then
guaranteed by Theorem~\ref{th:smooth}. We rewrite
\eqref{eq:Fuchsian-low} as
\begin{subequations}\label{eq:symmhyp-low}
\begin{align}
  \label{eq:symmhyp-low-DU0}
  DU_0 =&\; U_1, \\
  \begin{split}\label{eq:symmhyp-low-DU1}
    (1-\tau^2) DU_1 
    =& -(1-\tau^2) \bigl[ 2\epsilon U_1 + \epsilon^2 U_0 \bigr]
    + \tau^{2-\epsilon} \bigl[ k + (\ln\tau)\lap{k} + \lap\varphi \bigr]
    + \ignorewidth{ \tau^2 (U_1 + \epsilon U_0) } \\% ignorewidth
    &+ \tau \bigl[
              (1-x^2)U_{2x} - xy(U_{2y}+U_{3x}) + (1-y^2) U_{3y}
              - 2x U_2 - 2y U_3
            \bigr] \\
  &- e^{-2\varphi-2\tau^{\epsilon}U_0} 
      \Bigl\{
        (1-\tau^2)\tau^{2k-\epsilon} \bigl[V_1 + 2k(\psi+V_0)\bigr]^2
        - \tau^{2-2k-\epsilon}(\grad\omega_0)^2 \\
  &\phantom{- e^{-2\varphi-2\tau^{\epsilon}U_0} \Bigl\{}
        - 2\tau^{1-\epsilon} \grad\omega_0 \cdot
            \bigl[
              \Vbar + \tau\grad\psi + 2\tau(\ln\tau)(\psi+V_0)\grad{k}
            \bigr] \\
  &\phantom{- e^{-2\varphi-2\tau^{\epsilon}U_0} \Bigl\{}
        - \tau^{2k-\epsilon} 
            \bigl[
              \Vbar + \tau\grad\psi + 2\tau(\ln\tau)(\psi+V_0)\grad{k}
            \bigr]^2
      \Bigr\},
  \end{split} \\
  \label{eq:symmhyp-low-DU2}
  (1-x^2) DU_2\;-& \; xy DU_3 
    = (1-x^2) U_2 - xy U_3 + \tau(1-x^2) U_{1x} - \tau xy U_{1y}, \\
  \label{eq:symmhyp-low-DU3}
  (1-y^2) DU_3 \;-&\; xy DU_2
    = (1-y^2) U_3 - xy U_2 + \tau(1-y^2) U_{1y} - \tau xy U_{1x}, 
    \displaybreak[0]\\
  DV_0 =&\; V_1, \\
  \begin{split}\label{eq:symmhyp-low-DV1}
    (1-\tau^2) DV_1 =& -2k (1-\tau^2) V_1 
    + 2(1-\tau^2)\tau^\epsilon 
        (U_1+\epsilon U_0) \bigl( V_1+2k(\psi+V_0) \bigr) \\
    &+ \tau^2 \ignorewidth{ \bigl[% begin ignorewidth
                V_1 + 2k(\psi+V_0)
                + 2(\ln\tau)(\psi+V_0)\lap{k}
                + 2(\ln\tau)\grad{k}\cdot\grad\psi + \lap\psi
              \bigr] } \\% end ignorewidth
    &+ 2\tau(\ln\tau) \grad{k}\cdot\Vbar + \tau^{2-2k}\lap\omega_0 \\
    &+ \tau \bigl[
              (1-x^2)V_{2x} - xy(V_{2y}+V_{3x}) + (1-y^2) V_{3y}
              - 2x V_2 - 2y V_3
            \bigr] \\
    &- 2\tau^{2-2k}\grad\omega_0\cdot
        \bigl[ 
          (\ln\tau)\grad{k} + \grad\varphi
        \bigr]
     - 2\tau^{1+\epsilon-2k}\grad\omega_0\cdot\Ubar \\
    &- 2\tau^\epsilon
        \bigl[ \tau^{1-\epsilon}\grad\varphi + \Ubar \bigr]
        \cdot 
        \bigl[
          2\tau(\ln\tau)(\psi+V_0)\grad{k} + \tau\grad\psi + \Vbar
        \bigr],
  \end{split} \\
  (1-x^2) DV_2\;-& \; xy DV_3 
    = (1-x^2) V_2 - xy V_3 + \tau(1-x^2) V_{1x} - \tau xy V_{1y}, \\
  (1-y^2) DV_3 \;-&\; xy DV_2
    = (1-y^2) V_3 - xy V_2 + \tau(1-y^2) V_{1y} - \tau xy V_{1x}.
\end{align}
\end{subequations}

Because of the factor $\tau^{1+\epsilon-2k}$, we must restrict $k$ to
the case $0<k<3/4$.  After an appropriate rescaling of the time
variable, the system is clearly a regular symmetric hyperbolic system
of the form \eqref{eq:symmhyp}. Forming the corresponding positive
definite system (where $i=3$ suffices to make the coefficient of the
singular term positive definite) and applying Theorem~\ref{th:smooth}
then gives the existence of the desired solutions.

Note that there is a slight difficulty in showing that
\eqref{eq:symmhyp-low} is equivalent to \eqref{eq:Fuchsian-low}.  For
example, from \eqref{eq:symmhyp-low-DU2}, \eqref{eq:symmhyp-low-DU3}
and \eqref{eq:symmhyp-low-DU0} it follows that
$D(U_2-\tau\d_x{U_0})=U_2-\tau\d_x{U_0}$, and since $U_2$ and $U_0$
vanish at $\tau=0$ we must have $U_2-\tau\d_x{U_0}=c(x)\tau$ for some
function $c(x)$.  But there seems to be no direct way of showing that
$c(x)\equiv0$ follows from \eqref{eq:symmhyp-low}.  This must be the
case however since we obtain the solution as the limit of a sequence
of analytic solutions with this property.

% Note that there is a slight difficulty in showing that
% \eqref{eq:symmhyp-low} is equivalent to \eqref{eq:Fuchsian-low}. For
% example, from \eqref{eq:symmhyp-low-DU2}, \eqref{eq:symmhyp-low-DU3}
% and \eqref{eq:symmhyp-low-DU0} it follows that
% $D(U_2-\tau\d_x{U_0})=U_2-\tau\d_x{U_0}$, and since $U_2$ and $U_0$
% vanish at $\tau=0$ we must have $U_2-\tau\d_x{U_0}=c(x)\tau$ for some
% function $c(x)$. But it seems to be difficult to show that $c(x)=0$
% follows from \eqref{eq:symmhyp-low}. This problem can be circumvented
% by instead of trying to show equivalence of \eqref{eq:Fuchsian-low}
% and the positive definite system \eqref{eq:symmhyp-diff} using
% \eqref{eq:symmhyp-low}, we use a Fuchsian positive definite system
% obtained from \eqref{eq:Fuchsian-low} in the same way as
% \eqref{eq:symmhyp-diff} was obtained from \eqref{eq:symmhyp-low}. The
% Fuchsian positive definite system will then be equal to
% \eqref{eq:symmhyp-diff} except for the linear transformation of
% $(DU_2,DU_3)$ and $(DV_2,DV_3)$ and replacement of $DU_2$, $DU_3$,
% $DV_2$ and $DV_3$ according to their definitions \eqref{eq:UVdef} in
% some places. The equivalence with the original Fuchsian system
% \eqref{eq:Fuchsian-low} is straightforward, and there is no problem
% with the equivalence with \eqref{eq:symmhyp-diff} since now the
% coefficient of the singular term is positive definite.

We summarise this section in the following theorem.
\begin{theorem}\label{th:smooth-low}
  If $k$, $\varphi$, $\omega_0$ and $\psi$ are smooth functions of
  $\theta$ and $0<k<3/4$ for all $\theta$, there exists a solution of
  Einstein's equations \eqref{eq:ee-Gowdy} in a neighbourhood $\U$ of
  $\theta=0$ of the form \eqref{eq:expansion-low} where
  $2k-1<\epsilon<\min\{2k,2-2k\}$ and $u$ and $v$ are regular and tend
  to $0$ as $t\to0$. Given the form of the expansion and a choice of
  $\epsilon$, the solution is unique.
\end{theorem}

% % % % % % % % % % % % % % % % % % % % % % % % % % % % % % % % % % % % % % % %

\subsection{Negative velocity, smooth case}
\label{sec:smooth-neg}

Since exponents of $\tau$ involving $-2k$ only appear in terms
containing $\grad\omega_0$, the argument in the previous section
applies immediately to the negative velocity case with $k>0$ and
constant $\omega_0$.
\begin{theorem}\label{th:smooth-high}
  If $k$, $\varphi$ and $\psi$ are smooth functions of $\theta$ such
  that $k>0$ for all $\theta$ and $\omega_0$ is a constant, there
  exists a solution of Einstein's equations \eqref{eq:ee-Gowdy} in a
  neighbourhood $\U$ of $\theta=0$ of the form
  \eqref{eq:expansion-low} where $\epsilon<2k$ and $u$ and $v$ are
  regular and tend to $0$ as $t\to0$. Given the form of the expansion
  and a choice of $\epsilon$, the solution is unique.
\end{theorem}

% % % % % % % % % % % % % % % % % % % % % % % % % % % % % % % % % % % % % % % %

\subsection{High velocity, smooth case}
\label{sec:smooth-high}

The calculations for the high velocity case are similar to the low
velocity case and will be omitted.  We have the following existence
result.
\begin{theorem}\label{th:smooth-neg}
  If $k$, $\varphi$ and $\omega_0$ are smooth functions of $\theta$
  and $k<1/2$ for all $\theta$, there exists a solution of Einstein's
  equations \eqref{eq:ee-Gowdy} in a neighbourhood $\U$ of $\theta=0$
  of the form \eqref{eq:expansion-neg} where
  $\max\{0,2k\}<\epsilon<\min\{2,2-2k\}$ and $u$ and $v$ are regular
  and tend to $0$ as $t\to0$. Given the form of the expansion and a
  choice of $\epsilon$, the solution is unique.
\end{theorem}

% % % % % % % % % % % % % % % % % % % % % % % % % % % % % % % % % % % % % % % %

\subsection{Intermediate velocity, smooth case}
\label{sec:smooth-int}

It remains to treat the case when $3/4\le k<1$.  The idea is to
include one more term in the expansion \eqref{eq:expansion-low-Z} of
$Z$,
\begin{equation}\label{eq:expansion-int}
  Z(\tau,\theta) = k(\theta)\ln\tau + \varphi(\theta) 
    + \alpha(\theta)\tau^{2-2k} + \tau^{2-2k+\epsilon} u(\tau,\theta),
\end{equation}
where $\alpha$ is to be chosen such as to eliminate the leading order
terms in the equations.  We keep the original expansion for $\omega$
as in \eqref{eq:expansion-low-omega}.

The resulting system is similar to \eqref{eq:symmhyp-low}, but with
$DU_1$ and $DU_2$ given by
\begin{subequations}\label{eq:symmhyp-int}
\begin{align}
  \begin{split}\label{eq:symmhyp-int-DU1}
    (1-\tau^2) DU_1 
    =& -(1-\tau^2)
          \bigl[ 
            (4-4k+2\epsilon)U_1 + (2-2k+\epsilon)^2 U_0
          \bigr]
     - 4\tau(\ln\tau)\grad{k}\cdot\Ubar \\
    &+ \tau^{2k-\epsilon}
         \bigl[ k + (\ln\tau)\lap{k} + \lap\varphi \bigr]
     + \tau^{2-\epsilon}(4k^2-10k+6)\alpha \\
    &+ \tau^{2-\epsilon}
         \bigl[
           \lap\alpha
           - 2(\ln\tau)(\alpha\lap{k} + 2\grad\alpha\cdot\grad{k})
           + 4(\ln\tau)^2\alpha(\grad{k})^2
         \bigr] \\
    &+ \tau^2 \bigl[ 
                U_1 + (2-2k+\epsilon)U_0 - 2(\ln\tau)U_0\lap{k}
                + 4(\ln\tau)^2 U_0 (\grad{k})^2
              \bigr] \\
    &+ \tau \bigl[
              (1-x^2)U_{2x} - xy(U_{2y}+U_{3x}) + (1-y^2) U_{3y}
              - 2x U_2 - 2y U_3
            \bigr] \\
  &- \exp(-2\varphi-2\alpha\tau^{2-2k} - 2\tau^{2-2k+\epsilon}U_0) \\
  &\quad\times\Bigl\{
       (1-\tau^2)\tau^{4k-2-\epsilon} \bigl[V_1 + 2k(\psi+V_0)\bigr]^2 \\
  &\phantom{\quad\times\Bigl\{}
       - 2\tau^{2k-1-\epsilon} \grad\omega_0 \cdot
           \bigl[
             \Vbar + \tau\grad\psi + 2\tau(\ln\tau)(\psi+V_0)\grad{k}
           \bigr] \\
  &\phantom{\quad\times\Bigl\{}
       - \tau^{4k-2-\epsilon} 
           \bigl[
             \Vbar + \tau\grad\psi + 2\tau(\ln\tau)(\psi+V_0)\grad{k}
           \bigr]^2
     \Bigr\} \\
  &+ \tau^{-\epsilon}
       \bigl[
         \exp(-2\varphi-2\alpha\tau^{2-2k} - 2\tau^{2-2k+\epsilon}U_0)
           (\grad\omega_0)^2
         - (2-2k)^2 \alpha
       \bigr],
  \end{split} \displaybreak[0]\\
  \begin{split}\label{eq:symmhyp-int-DV1}
     (1-\tau^2) DV_1 =& -2k (1-\tau^2) V_1
     + 2\tau(\ln\tau) \grad{k}\cdot\Vbar \\
    &+ 2(1-\tau^2)\tau^{2-2k+\epsilon}
        \bigl( U_1+(2-2k+\epsilon)U_0 \bigr)
        \bigl( V_1+2k(\psi+V_0) \bigr) \\
    &+ \tau^2 \bigl[
                V_1 + 2k(\psi+V_0)
                + 2(\ln\tau)(\psi+V_0)\lap{k}
                + 2(\ln\tau)\grad{k}\cdot\grad\psi + \lap\psi
              \ignorewidth{\bigr]} \\% ignorewidth
    &+ \tau^{2-2k}
         \bigl[
           \lap\omega_0 
           + 2(1-\tau^2)(2-2k)\alpha\bigl(V_1+2k(\psi+V_0)\bigr)
         \bigr] \\
    &+ \tau \bigl[
              (1-x^2)V_{2x} - xy(V_{2y}+V_{3x}) + (1-y^2) V_{3y}
              - 2x V_2 - 2y V_3
            \bigr] \\
    &- 2\tau^{2-2k}\grad\omega_0\cdot
        \bigl[ 
          (\ln\tau)\grad{k} + \grad\varphi
        \bigr]
     - 2\tau^{3-4k+\epsilon}\grad\omega_0\cdot\Ubar \\
    &- 2\tau^{4-4k}\grad\omega_0\cdot
        \bigl[ 
          \grad\alpha - 2(\ln\tau)(\alpha+\tau^{\epsilon}U_0)\grad{k}
        \bigr] \\
    &- 2\tau^{2-2k}
        \bigl[
          \tau^{2k-1}\grad\varphi + \tau\grad\alpha
          - 2\tau(\ln\tau)(\alpha+\tau^{\epsilon}U_0)\grad{k}
          + \Ubar
        \bigr] \\
    &\qquad\cdot
        \bigl[ 
          2\tau(\ln\tau)(\psi+V_0)\grad{k} + \tau\grad\psi + \Vbar
        \bigr],
  \end{split}
\end{align}
\end{subequations}

We choose $\alpha:=(2-2k)^{-2}(\grad\omega_0)^2\exp(-2\varphi)$ to
cancel the $\tau^{-\epsilon}$ factor in the last term on the right of
\eqref{eq:symmhyp-int-DU1}.  That term will then be of order
$\tau^{2-2k-\epsilon}$, so the order of the $(\grad\omega_0)^2$ term
is unchanged from the previous case \eqref{eq:symmhyp-low-DU1}.

The problematic exponents of $\tau$ are now $2-2k-\epsilon$,
$3-4k+\epsilon$ and $2k-1-\epsilon$. We have to choose the number
$\epsilon>0$ such that $4k-3<\epsilon<2k-1$, but we also need to
ensure that $2-2k-\epsilon>0$, which is compatible with
$4k-3<\epsilon$ if and only if $1/2<k<5/6$. In that case, we may apply
the arguments of section~\ref{sec:regular} and \ref{sec:smooth-exist}
to show that a regular solution exists.

Contrary to what was claimed in \cite{Rendall:Gowdy-T3-smooth}, it is
not possible to cover the whole range $1/2<k<1$ since we can only
choose $\alpha$ such that the last term in \eqref{eq:symmhyp-int-DU1}
vanishes to first order in $\tau$. In equation~(27) of
\cite{Rendall:Gowdy-T3-smooth} the corresponding higher order terms
have been left out.

We can however cover small intervals of $k$ closer to $1$ by repeating
the method above. Replacing the expansion \eqref{eq:expansion-low-Z}
by \eqref{eq:expansion-int} is equivalent to performing the
transformation $u\mapsto\alpha\tau^{2-2k-\epsilon}+\tau^{2-2k}u$.
Repeating this transformation $i$ times, where each $\alpha$ is
chosen such that the leading order coefficient of $(\grad\omega_0)^2$
is cancelled at each stage, gives a system with positive powers of
$\tau$ if and only if
\begin{equation}\label{eq:k-interval-i}
  1-\frac{1}{2i} < k < 1-\frac{1}{2i+4}
\end{equation}
and
\begin{equation}\label{eq:eps-interval-i}
  1 - 2(i+1)(1-k) < \epsilon < \min\{2-2k, 1-2i(1-k)\}.
\end{equation}
The interval $(1/2,1)$ is covered by the infinite sequence of
intervals $(1-1/2i,1-1/(2i+4))$. The expansion of $Z$ for a given $i$
is
\begin{equation}\label{eq:expansion-int-i}
  Z(\tau,\theta) = k(\theta)\ln\tau + \varphi(\theta) 
    + \sum_{j=1}^i \alpha_j(\theta)\tau^{(2-2k)j}
    + \tau^{(2-2k)i+\epsilon} u(\tau,\theta).
\end{equation}

Note that as $k$ tends to $1$, we have to include an increasing number
of terms in the expansion \eqref{eq:expansion-int-i} of $Z$.  This
might seem to be contradictory since we keep the original expansion of
$\omega$, and the higher order terms of $\omega$ should affect $Z$ at
some finite order.  But for any given $k$, the number of terms in
\eqref{eq:expansion-int-i} is finite since $i$ is bounded above by
\eqref{eq:k-interval-i}. Also, the exponent of $\tau$ in the terms
containing $\alpha_j$ is always between $0$ and $1$, so higher order
contributions are still encoded in $u$.

The above discussion motivates the following theorem.
\begin{theorem}\label{th:smooth-int}
  If $k$, $\varphi$, $\omega_0$ and $\psi$ are smooth functions of
  $\theta$ and $k$ satisfies \eqref{eq:k-interval-i} for all $\theta$
  and some natural number $i$, there exists a solution of Einstein's
  equations \eqref{eq:ee-Gowdy} in a neighbourhood $\U$ of $\theta=0$
  of the form \eqref{eq:expansion-int-i} and
  \eqref{eq:expansion-low-omega} where $\epsilon$ satisfies
  \eqref{eq:eps-interval-i} and $u$ and $v$ are regular and tend to
  $0$ as $t\to0$. Given the form of the expansion and a choice of
  $\epsilon$, the solution is unique.
\end{theorem}

%%%%%%%%%%%%%%%%%%%%%%%%%%%%%%%%%%%%%%%%%%%%%%%%%%%%%%%%%%%%%%%%%%%%%%%%%%%%%%%

\section{Discussion}
\label{sec:discussion}

We have constructed families of solutions to the evolution equations
of Gowdy spacetimes with $\sphere^2\times\sphere^1$ or $\sphere^3$
spatial topology.  When the asymptotic velocity is between $0$ and $1$
($0<k<1$) we obtain solutions depending on four free functions, which
is the same number as in the general solution, while outside that
range the solutions include only three free functions. The solutions
are asymptotically velocity dominated by construction. Unfortunately,
for regular asymptotically velocity dominated solutions of the full
Einstein equations the asymptotic velocity must be $-1$ or $3$
($k=\pm2$) at the axes.  This can also be checked directly by
inserting the asymptotic expansions of the solutions into the
constraints \eqref{eq:ec-Ernst}.  Since both analytical and numerical
arguments indicate that the velocity is between $0$ and $1$ for
$\torus^3$ Gowdy, and these results may be applied on sets not
containing the axes, it seems that there will be spikes at the axes in
general situations.

There are two possible interpretations.  Firstly, there is the
possibility of false spikes, which result from an inappropriate choice
of parametrisation of the metric.  This is also what makes our
solutions different from the numerical solutions
\cite{Garfinkle:Gowdy-S2xS1-num}.  The parametrisation problem may be
traced to different choices of parametrisations of the hyperbolic
plane.  It might be possible to choose this parametrisation
differently such as to avoid coordinate singularities.  This is of
equal importance in the study of the dynamics of $\torus^3$ Gowdy, in
particular when trying to verify the numerical observation that the
velocity is eventually driven below $1$ in a mathematically rigorous
way.

Secondly, there is the possibility of true spikes at the axes.  In
\cite{Rendall-Weaver:manufacture}, $\torus^3$ Gowdy solutions with
spikes are constructed by applying suitable transformations to a given
solution with velocity between $0$ and $1$.  At first sight it seems
possible to do something similar for the other topologies.  The
transformations used in \cite{Rendall-Weaver:manufacture} are
inversion, which interchanges the Killing vectors, and a
Gowdy-to-Ernst transformation.  A combination of these transformations
preserve the evolution equations in the $\sphere^2\times\sphere^1$ and
$\sphere^3$ cases as well.  In fact, the equations for $Z$ and
$\omega$ are the same as the equations for $-P$ and $Q$ (in the
notation of section~\ref{sec:param}).  But $P$ is singular at the
axes, so this method of constructing solutions with spikes does not
work in our case.

Finally, some remarks on the asymptotic behaviour of the curvature is
in order.  Using the asymptotic expansions for our solutions,
\eqref{eq:expansion-low} or \eqref{eq:expansion-neg}, it is
straightforward to calculate that a necessary condition for the
Kretschmann scalar $R_{ijkl}R^{ijkl}$ to be bounded is that the
asymptotic velocity is $\pm1$ ($k=0$ or $k=2$), which is in agreement
with the result for the polarised models described in
section~\ref{sec:polarsol}.  All of the black hole solutions mentioned
in section~\ref{sec:BHsol} are of course extendible through the
horizon, and the asymptotic velocity is indeed $\pm1$ there.  While
such non-generic solutions can be ignored from the cosmological point
of view, they might be interesting by analogue to black hole
interiors. One interesting question that remains to be answered is
under which circumstances general Gowdy models can be extended through
a compact Cauchy horizon. In particular, is analyticity necessary as
in the polarised case \cite{Chrusciel-Isenberg-Moncrief:Gowdy-SCC}?

To conclude, it seems that for more progress on Gowdy
$\sphere^2\times\sphere^1$ and $\sphere^3$ spacetimes to be made, a
better understanding and handling of the spikes are needed.  If this
can be done for $\torus^3$ Gowdy, it should be possible to adapt the
techniques to the other topologies using some of the arguments of this
paper.

%%%%%%%%%%%%%%%%%%%%%%%%%%%%%%%%%%%%%%%%%%%%%%%%%%%%%%%%%%%%%%%%%%%%%%%%%%%%%%%

\section*{Acknowledgements}

I thank Alan Rendall for helpful discussions and suggestions.

%%%%%%%%%%%%%%%%%%%%%%%%%%%%%%%%%%%%%%%%%%%%%%%%%%%%%%%%%%%%%%%%%%%%%%%%%%%%%%%

\providecommand{\bysame}{\leavevmode\hbox to3em{\hrulefill}\thinspace}


\begin{thebibliography}{10}

\bibitem{Andersson-Rendall:quiescent-sings}
L.~Andersson and A.~D. Rendall, \emph{Quiescent cosmological singularities},
  Comm. Math. Phys. \textbf{218} (2001), 479--511.

\bibitem{Anguige:isotropic-III}
K.~Anguige, \emph{Isotropic cosmological singularities {III}. {T}he {Cauchy}
  problem for the inhomogeneous conformal {Einstein-Vlasov} equations}, Ann.
  Physics \textbf{282} (2000), 395--419.

\bibitem{Anguige-Tod:isotropic-I}
K.~Anguige and K.~P. Tod, \emph{Isotropic cosmological singularities {I}.
  {P}olytropic perfect fluid spacetimes}, Ann. Physics \textbf{276} (1999),
  no.~2, 257--293.

\bibitem{Anguige-Tod:isotropic-II}
\bysame, \emph{Isotropic cosmological singularities {II}. {T}he
  {Einstein-Vlasov} system}, Ann. Physics \textbf{276} (1999), no.~2, 294--320.

\bibitem{Anguige:Gowdy-PF-Kasner}
K.~Anguige, \emph{A class of perfect-fluid cosmologies with polarised {Gowdy}
  symmetry and a {Kasner}-like singularity}, preprint, gr-qc/0005086, 2000.

\bibitem{Anguige:plane-PF-Kasner}
\bysame, \emph{A class of plane symmetric perfect-fluid cosmologies with a
  {Kasner}-like singularity}, Class. Quantum Grav. \textbf{17} (2000),
  2117--2128.

\bibitem{BKL:general-time-sing}
V.~A. Belinskii, I.~M. Khalatnikov, and E.~M. Lifshitz, \emph{A general
  solution of the {Einstein} equations with a time singularity}, Ann. Physics
  \textbf{31} (1982), 639--667.

\bibitem{Berger-Garfinkle:Gowdy-T3-phenom}
B.~K. Berger and D.~Garfinkle, \emph{Phenomenology of the {Gowdy} universe on
  {$T^3\times R$}}, Phys. Rev. D (3) \textbf{57} (1998), no.~8, 4767--4777.

\bibitem{Berger-Moncrief:Gowdy-T3-num}
B.~K. Berger and V.~Moncrief, \emph{Numerical investigation of cosmological
  singularities}, Phys. Rev. D (3) \textbf{48} (1993), no.~10, 4676--4687.

\bibitem{Chrusciel:U1xU1}
P.~T. Chru\'sciel, \emph{On space-times with {$U(1)\times U(1)$} symmetric
  compact {Cauchy} surfaces}, Ann. Physics \textbf{202} (1990), 100--150.

\bibitem{Chrusciel:uniqueness-II}
\bysame, \emph{On uniqueness in the large of solutions of {Einstein's}
  equations (``strong cosmic censorship'')}, Proc. Centre Math. Appl., vol.~27,
  Austral. Nat. Univ., 1991.

\bibitem{Chrusciel-Isenberg-Moncrief:Gowdy-SCC}
P.~T. Chru\'sciel, J.~Isenberg, and V.~Moncrief, \emph{Strong cosmic censorship
  in polarised {Gowdy} spacetimes}, Class. Quantum Grav. \textbf{7} (1990),
  1671--1680.

\bibitem{Claudel:Cauchy}
C.~M. Claudel, \emph{The cauchy problem for quasi-linear hyperbolic evolution
  problems with a singularity in the time}, Proc. Roy. Soc. London Ser. A
  \textbf{454} (1998), no.~1972, 1073--1107.

\bibitem{Eardley-Liang-Sachs:VTD-dust}
D.~Eardley, E.~Liang, and R.~Sachs, \emph{Velocity-dominated singularities in
  irrotational dust cosmologies}, J. Math. Phys. \textbf{13} (1972), no.~1,
  99--107.

\bibitem{Garfinkle:Gowdy-S2xS1-num}
D.~Garfinkle, \emph{Numerical simulations of {Gowdy} spacetimes on {$S^2\times
  S^1\times R$}}, Phys. Rev. D (3) \textbf{60} (1999), 104010--1--6.

\bibitem{Geroch-Hartle:distorted-BH}
R.~Geroch and J.~B. Hartle, \emph{Distorted black holes}, J. Math. Phys.
  \textbf{23} (1982), no.~4, 680--692.

\bibitem{Gowdy:Gowdy}
R.~H. Gowdy, \emph{Vacuum spacetimes with two-parameter spacelike isometry
  groups and compact invariant hypersurfaces: Topologies and boundary
  conditions}, Ann. Physics \textbf{83} (1974), 203--241.

\bibitem{Griffiths-Alekseev:unpolar-Gowdy}
J.~B. Griffiths and G.~A. Alekseev, \emph{Some unpolarized {G}owdy cosmologies
  and noncolinear colliding plane wave spacetimes}, Int. J. Modern Phys. D
  \textbf{7} (1998), no.~2, 237--247.

\bibitem{Grubisic-Moncrief:Gowdy-T3-asymptotic}
B.~Grubi\v{s}i\'c and V.~Moncrief, \emph{Asymptotic behavior of the {$T^3\times
  R$} {Gowdy} space-times}, Phys. Rev. D (3) \textbf{47} (1993), no.~6,
  2371--2382.

\bibitem{Hanquin-Demaret:Gowdy-S2xS1+S3}
J.-L. Hanquin and J.~Demaret, \emph{Gowdy ${S}^1\otimes{S}^2$ and ${S}^3$
  inhomogeneous cosmological models}, J. Phys. A \textbf{16} (1983), L5--L10.

\bibitem{Hawking-Ellis}
S.~W. Hawking and G.~F.~R. Ellis, \emph{The large scale structure of
  space-time}, Cambridge Univ. Press, Cambridge, 1973.

\bibitem{Hern-Stewart}
S.~D. Hern and J.~M. Stewart, \emph{The {Gowdy} {T-3} cosmologies revisited},
  Class. Quantum Grav. \textbf{15} (1998), no.~6, 1581--1593.

\bibitem{Hern:num-inhom}
S.~D. Hern, \emph{Numerical relativity and inhomogeneous cosmologies}, Ph.D.
  thesis, University of Cambridge, 1999.

\bibitem{Isenberg-Kichenassamy:Gowdy-T2-polar}
J.~Isenberg and S.~Kichenassamy, \emph{Asymptotic behavior in polarized
  {$T^2$}-symmetric vacuum space-times}, J. Math. Phys. \textbf{40} (1999),
  no.~1, 340--352.

\bibitem{Isenberg-Moncrief:Gowdy-polar-asymptotic}
J.~Isenberg and V.~Moncrief, \emph{Asymptotic behavior of the gravitational
  field and the nature of singularities in {Gowdy} spacetimes}, Ann. Physics
  \textbf{199} (1990), 84--122.

\bibitem{Kichenassamy:Fuchsian-blow-up}
S.~Kichenassamy, \emph{{Fuchsian} equations in {Sobolev} spaces and blow-up},
  J. Diff. Eqns \textbf{125} (1996), 299--327.

\bibitem{Kichenassamy:wave-eq}
\bysame, \emph{Nonlinear wave equations}, Marcel Dekker, New York, 1996.

\bibitem{Kichen-Rendall:Gowdy-T3-analytic}
S.~Kichenassamy and A.~D. Rendall, \emph{Analytic description of singularities
  in general relativity}, Class. Quantum Grav. \textbf{15} (1998), 1339--1355.

\bibitem{Kramer-Neugebauer:axialsymm}
D.~Kramer and G.~Neugebauer, \emph{Zu axialsymmetrischen station\"aren
  {L}\"osungen der {E}insteinschen {F}eldgleichungen f\"ur das {V}akuum}, Comm.
  Math. Phys. \textbf{10} (1968), 132--139.

\bibitem{Moncrief:U1-Cauchy-horizons}
V.~Moncrief, \emph{Neighborhoods of {Cauchy} horizons in cosmological
  spacetimes with one {Killing} field}, Ann. Physics \textbf{141} (1982),
  83--103.

\bibitem{Mostert:compact-action}
P.~S. Mostert, \emph{On a compact {Lie} group acting on a manifold}, Ann. of
  Math. \textbf{65} (1956), no.~3, 447--455.

\bibitem{Narita-Torii-Maeda:Gowdy-string}
M.~Narita, T.~Torii, and K.~Maeda, \emph{Asymptotic singular behaviour of
  {Gowdy} spacetimes in string theory}, Class. Quantum Grav. \textbf{17}
  (2000), 4597--4613.

\bibitem{Newman:conformal-sing}
R.~P. A.~C. Newman, \emph{On the structure of conformal singularities in
  classical general relativity}, Proc. Roy. Soc. London Ser. A \textbf{443}
  (1993), 473--492.

\bibitem{Newman:conformal-sing-II}
\bysame, \emph{On the structure of conformal singularities in classical general
  relativity. {II} evolution equations and a conjecture of {K}. {P}. {Tod}},
  Proc. Roy. Soc. London Ser. A \textbf{443} (1993), 493--515.

\bibitem{Obregon-Ryan:Gowdy-S2xS1-Kerr}
O.~Obreg\'on and J.~Michael P.~Ryan, \emph{A family of exact solutions for
  unpolarized {Gowdy} models}, preprint, gr-qc/9810068, 1998.

\bibitem{Rendall-Tod:EV-LRS}
A.~D. Rendall and K.~P. Tod, \emph{Dynamics of spatially homogeneous solutions
  of the {E}instein-{V}lasov equations which are locally rotationally
  symmetric}, Class. Quantum Grav. \textbf{16} (1999), 1705--1726.

\bibitem{Rendall-Uggla:EV-LRS}
A.~D. Rendall and C.~Uggla, \emph{Dynamics of spatially homogeneous locally
  rotationally symmetric solutions of the {E}instein-{V}lasov equations},
  Class. Quantum Grav. \textbf{17} (2000), 4697--4713.

\bibitem{Rendall:Gowdy-T3-smooth}
A.~D. Rendall, \emph{{Fuchsian} analysis of singularities in {Gowdy} spacetimes
  beyond analyticity}, Class. Quantum Grav. \textbf{17} (2000), 3305--3316.

\bibitem{Rendall-Weaver:manufacture}
A.~D. Rendall and M.~Weaver, \emph{Manufacture of {G}owdy spacetimes with
  spikes}, Class. Quantum Grav. \textbf{18} (2001), 2959--2976.

\bibitem{Ringstrom:curvature-blow-up}
H.~Ringstr\"om, \emph{Curvature blow-up in {Bianchi} {VIII} and {IX} vacuum
  spacetimes}, Class. Quantum Grav. \textbf{17} (2000), 713--731.

\bibitem{Ringstrom:BianchiIX}
\bysame, \emph{The {B}ianchi {IX} attractor}, Ann. Henri Poincar\'e \textbf{2}
  (2001), 405--500.

\bibitem{Tanimoto:local-U1xU1}
M.~Tanimoto, \emph{Locally ${U}^1\times{U}^1$ symmetric cosmological models:
  Topology and dynamics}, Class. Quantum Grav. \textbf{18} (2001), 479--507.

\bibitem{Weaver:magn-Bianchi-IX0}
M.~Weaver, \emph{Dynamics of magnetic {Bianchi} {IV}$_0$ cosmologies}, Class.
  Quantum Grav. \textbf{17} (2000), 421--434.

\end{thebibliography}
\end{document}